\documentclass[onecolumn,useAMS,usenatbib]{mn2e}
\usepackage{graphicx}
\usepackage{bm}
\usepackage{amsfonts}
\usepackage{latexsym}
\usepackage[latin1]{inputenc}
\usepackage{amsmath}
\usepackage{epsfig}
\usepackage{amsbsy}
\usepackage{mnfixes}

\newcommand{\mtb}[1]{\mathbf{#1}}

\newcommand{\n}{\mathrm n}
\newcommand{\p}{\mathrm p}
\newcommand{\e}{\mathrm e}
\newcommand{\x}{\mathrm x}
\newcommand{\f}{\mathrm f}

\def\jnl@style{\rm}
\def\aaref@jnl#1{{\jnl@style#1}}

\def\aaref@jnl#1{{\jnl@style#1}}

\def\aj{\aaref@jnl{AJ}}                   
\def\apj{\aaref@jnl{ApJ}}                 
\def\apjl{\aaref@jnl{ApJ}}                
\def\apjs{\aaref@jnl{ApJS}}               
\def\apss{\aaref@jnl{Ap\&SS}}             
\def\aap{\aaref@jnl{A\&A}}                
\def\aapr{\aaref@jnl{A\&A~Rev.}}          
\def\aaps{\aaref@jnl{A\&AS}}              
\def\mnras{\aaref@jnl{MNRAS}}             
\def\prd{\aaref@jnl{Phys.~Rev.~D}}        
\def\prl{\aaref@jnl{Phys.~Rev.~Lett.}}    
\def\qjras{\aaref@jnl{QJRAS}}             
\def\skytel{\aaref@jnl{S\&T}}             
\def\ssr{\aaref@jnl{Space~Sci.~Rev.}}     
\def\zap{\aaref@jnl{ZAp}}                 
\def\nat{\aaref@jnl{Nature}}              
\def\aplett{\aaref@jnl{Astrophys.~Lett.}} 
\def\apspr{\aaref@jnl{Astrophys.~Space~Phys.~Res.}} 
\def\physrep{\aaref@jnl{Phys.~Rep.}}      
\def\physscr{\aaref@jnl{Phys.~Scr}}       

\title{Oscillations of rapidly rotating stratified neutron stars}

\author[A. Passamonti et al]
{A. Passamonti\thanks{E-mail:a.passamonti@soton.ac.uk} , B. Haskell, 
N. Andersson, D.I. Jones and I. Hawke \\ \\
School of Mathematics, University of Southampton, Southampton SO17 1BJ, UK}

\begin{document}

\date{\today}

\pagerange{\pageref{firstpage}--\pageref{lastpage}} \pubyear{}
\maketitle

\label{firstpage}


\begin{abstract}
We use time-evolutions of the linear perturbation equations to study
the oscillations of rapidly rotating neutrons stars. Our models
account for the buoyancy due to composition gradients and we study,
for the first time, the nature of the resultant g-modes in a fast
spinning star.  We provide detailed comparisons of non-stratified and
stratified models. This leads to an improved understanding of the
relationship between the inertial modes of a non-stratified star and
the g-modes of a stratified system. In particular, we demonstrate that
each g-mode becomes rotation-dominated, i.e. approaches a particular
inertial mode, as the rotation rate of the star is increased. We also
discuss issues relating to the gravitational-wave driven instability
of the various classes of oscillation modes.
\end{abstract}


\section{Introduction}

The oscillations of rotating neutron stars are of great interest in
astrophysics. Recent efforts to understand the various pulsation modes
of compact stars have to a large extent been motivated by
gravitational-wave astronomy~\citep{1996PhRvL..77.4134A,
1998MNRAS.299.1059A}.  Different modes of oscillation depend on
different pieces of internal physics and one may hope to use
observations to probe, for example, the composition of the
high-density region~\citep{2008GReGr..40..945F,2007GReGr..39.1323B}.
Work in this area, which has a long history \citep[for a review see
e.g.,][]{lrr-1999-2}, gained further momentum recently with the
relatively successful matching between observed quasiperiodic
oscillations in the tails of magnetar flares and calculated torsional
oscillations of the neutron star crust~\citep{1998ApJ...498L..45D,
2005ApJ...634L.153P, 2007AdSpR..40.1446W,2007MNRAS.374..256S}.  The
excitement following these, the first ever, observations of likely
neutron star vibrations is significant. It is clear that our models
need to be improved significantly if we expect to carry out a
quantitative astero-seismology programme for neutron stars, but the
prospects for improvements look good. After all, we already have a
reasonable understanding of the dynamics of the crust region
\citep[see][]{2006CQGra..23.5367C,2007MNRAS.374..256S} as well as the
superfluid components in the core of the star~\citep{lrr-2007-1}.

 The present work is motivated by the need to consider more realistic
models of rapidly rotating stars.  Our emphasis will be on the role of
the internal stratification associated with composition
variations. This leads to the presence of the so-called
g-modes~\citep{1992ApJ...395..240R}, and we want to investigate how
these modes are affected by fast rotation (the analogous problem for
thermal ocean g-modes of rapidly rotating neutron stars has already
been considered, see \citet*{1996ApJ...460..827B}).  In order to
determine the rotational effects on the oscillation spectrum we study
linear perturbations of axisymmetric stellar configurations, where the
Coriolis force and the centrifugal flattening of the star are included
in the background.  In this way, we go beyond the so-called slow
rotation approximation, where the rotation itself is treated
perturbatively and the mode-frequencies of non-rotating models are
determined by a perturbation expansion with respect to the angular
velocity of the star $\Omega$.
We are particularly interested in the relationship between the g-modes
of a stratified stellar model and the inertial
modes~\citep{1999ApJ...521..764L}, for which the Coriolis force is the
main restoring agent.  The inertial modes are important since they may
be driven unstable by the emission of gravitational waves. The
strongest instability is associated with the so-called
r-modes~\citep{2001IJMPD..10..381A,Andersson:2002ch}.  A key issue
concerns the amplitude at which an unstable mode saturates. Detailed
work has shown that an unstable r-mode saturates due to nonlinear
couplings to other inertial modes~\citep{2003ApJ...591.1129A}.
However, the mode saturation has only been considered for slowly
rotating, non-stratified models and it is important to ask to what
extent these results will change if the models are made more
realistic. 
 In stratified neutron stars, the g-modes tend to become rotationally
dominated beyond a certain rotation rate~\citep[see
e.g.,][]{2000ApJS..129..353Y, 2000A&A...354...86D} and may then be
called inertia-gravity modes~\citep{1989nos..book.....U}.  Effects due
to the stratification are thus weakened, which indicates that the
r-mode saturation estimates may still be reliable, at least for the
fastest spinning systems.

Improving and extending a numerical code that has been used to study
non-stratified stars~\citep*{2002MNRAS.334..933J}, we approach the
problem via time-evolutions of the equations that govern the linear
perturbations of a, potentially rapidly, rotating Newtonian star.
This strategy has the advantage that one does not have to deal with
the many rotationally coupled multipoles, that tend to make a
slow-rotation calculation less tractable.  On the other hand, a
numerical evolution is not expected to have the precision of a
frequency domain calculation. Neither should one expect it to yield
the complete spectrum of modes. After all, the simulation results
depend on the chosen initial data. In absence of a clear idea of the
nature of the various modes one does not know how to excite specific
oscillations. One way around this problem is to follow the
``recycling'' strategy developed by~\citet*{Stergioulas:2003ep} and
\citet*{Dimmelmeier:2005zk}. Another, more pragmatic, approach is to
simply study the modes that are excited by ``generic'' initial data.
This is the attitude that we adopt here.

\section{Rotating stellar models} \label{sec:Stmodel}

We want to investigate the effect that composition variations have on
the various oscillation modes of a rapidly rotating neutron star. To
do this, we consider an equilibrium configuration representing a
rapidly and uniformly rotating star containing three particle species.
We assume that matter is composed of degenerate neutrons $n$, protons
$p$ and electrons $e$. For simplicity, we further assume that neutrons
and protons are not superfluid. In essence, our model represents the
conditions in the outer core of a very young neutron star that has not
yet cooled below the temperature at which the baryons form superfluid
condensates (in practice with a core temperature significantly above
$10^9$~K).  We also do not consider the various exotic phases of
matter that may be present in the deep core of a realistic neutron
star model, e.g. hyperons or deconfined
quarks~\citep[see][]{2004Sci...304..536L,2007PhR...442..109L}.
Accounting for the presence of such phases would not be too difficult,
but at this stage our focus is on understanding the basic role of
composition variations.  We are already extending this work to account
for the presence of superfluid components and expect to report on
results in this direction soon.

The matter composition is represented by the three number densities
$n_\x$, where $\x$ is either $\n$, $\p$ or $\e$. Charge neutrality is
provided by the Coulomb interaction, that efficiently lock together
protons and electrons~\citep*[e.g.,][]{2005PhRvD..71h3001V}. Thus we have
$n_\p=n_\e$ and $\mtb{v}_\p=\mtb{v}_\e$, where $\mtb{v}_\x$ denotes the
velocity of each fluid component. In this model, the pressure $P$ is a
function of two variables, e.g. the neutron and proton number
densities or alternatively the total baryon number density
$n_\mathrm{b} =n_\n + n_\p$ and the proton fraction $x_\p = n_\p /
n_\mathrm{b}$. In the unperturbed stellar model, it is assumed that
the weak interaction processes:
\begin{equation}
n \longrightarrow p + e + \bar{\nu}_{\rm e} \, , \qquad \qquad p + e
\longrightarrow n + \nu_{\rm e} \, ,
\end{equation}
lead to $\beta$-equilibrium. The proton fraction is then only a
function of the baryon number density (or equivalently the baryon mass
density $\rho = m \left( n_\p + n_\n \right)$, where for simplicity
the proton and neutron masses have been assumed equal). Then the
stellar fluid can be described by a barotropic equation of state $P =
P\left( \rho \right)$. In this paper, we consider a simple polytropic
model
\begin{equation}
P = K \rho ^{\Gamma _{\beta}} \, ,
\end{equation}
where $K$ is constant and
\begin{equation}
\Gamma_{\beta} \equiv \frac{d \log P}{ d \log \rho}  \, , \label{eq:Gbdef}
\end{equation}
is the adiabatic index.

Once the equation of state is provided, rotating stellar
configurations can be determined by the Hachisu self-consistent field
method~\citep{1986ApJS...61..479H}. For this purpose we use the
numerical code described in~\citet{2002MNRAS.334..933J}. This code
generates axisymmetric equilibrium models by specifying the polytropic
index and the ratio of the polar to equatorial axes $R_p/R_{eq}$.  In
Table \ref{tab:back-models} we give some details relating to the
rotation rate, energies and masses of equilibrium configurations used
in this paper.

\begin{table}
\begin{center}
\caption{\label{tab:back-models} This table displays the main
quantities of our rotating equilibrium configurations. The stellar models are
described by a polytropic equation of state with adiabatic index $\Gamma_{\beta}
= 2$. In the first and second columns we show, respectively, the
ratio of polar to equatorial axes and the angular velocity of the
star. In the third column, the rotation rate is compared to the
Kepler velocity $\Omega_K$ that represents the mass shedding
limit. The ratio between the rotational kinetic energy and
gravitational potential energy $T/|W|$ and the stellar mass are given in the
fourth and fifth columns, respectively. All quantities are given in
dimensionless units, where $G$ is the gravitational constant, 
$\rho_c$ represents the central mass density and $R_{eq}$ is the equatorial radius.}
\begin{tabular}{c  c c c c  }
\hline
  $ R_p / R_{eq} $  &  $ \Omega / \sqrt{G\rho_c}$ & $\Omega / \Omega_K$  & $ T/|W| \times 10^{2}$
  & $ M / (\rho_c R_{eq}^3)$  \\
\hline
   1.000            &         0.000               &           0.000      &    0.000 &    1.273   \\
   0.996            &         0.084               &           0.116      &    0.096 &    1.270   \\
   0.992            &         0.119               &           0.164      &    0.192 &    1.261   \\
   0.983            &         0.167               &           0.230      &    0.385 &    1.248   \\
   0.950            &         0.287               &           0.396      &    1.170 &    1.197   \\
   0.933            &         0.331               &           0.456      &    1.570 &    1.171   \\
   0.900            &         0.403               &           0.556      &    2.380 &    1.118   \\
   0.850            &         0.488               &           0.673      &    3.640 &    1.038   \\
   0.800            &         0.556               &           0.767      &    4.930 &    0.956   \\
   0.750            &         0.612               &           0.844      &    6.250 &    0.869   \\
   0.700            &         0.658               &           0.907      &    7.570 &    0.779   \\
   0.650            &         0.693               &           0.956      &    8.820 &    0.684   \\
   0.600            &         0.717               &           0.989      &    9.860 &    0.579   \\
\hline   
\end{tabular}
\end{center}
\end{table}
\section{The perturbation problem} \label{sec:pert-eqs}

In a perturbed configuration, an oscillating fluid element may not
have time to reach $\beta$-equilibrium with the neighbouring
matter. In fact, for typical stellar oscillation frequencies the time
scale of the weak interaction processes is much too long for the
matter to equilibrate within one oscillation
period~\citep[e.g.,][]{1992ApJ...395..240R}.  In the limit of slow
reactions one can thus accurately assume that the composition of a
perturbed fluid element is frozen. This condition can be imposed by
letting the Lagrangian variation of the proton fraction vanish, i.e.,
\begin{equation}
\Delta x_{\p} = \delta x_{\p} + \mtb{\xi}  \cdot \nabla x_{\p} = 0  \, , \label{eq:Dxp}
\end{equation}
where $\mtb{\xi}$ denotes the Lagrangian displacement of a fluid
element.

The perturbed fluid is no longer described by the one-parameter
equation of state that determined the background model.  The perturbed
pressure now depends on both the proton fraction and the baryon
density,
\begin{equation}
 \delta P = \delta P \left( \rho, x_{\p} \right) \, . \label{eq:dPEoS}
\end{equation}
Given equation~(\ref{eq:Dxp}), the Lagrangian perturbation of pressure
is given by
\begin{equation}
\Delta P = \Gamma_{\rm f} \frac{P}{\rho} \, \Delta \rho \, , \label{eq:Dp}
\end{equation}
where
\begin{equation}
\Gamma_{\rm f} \equiv \left. \frac{\partial \log P }{\partial \log \rho}
\right|_{x_{\p}} \, , \label{eq:Gf}
\end{equation}
is the adiabatic index for a frozen composition. Alternatively, equation~(\ref{eq:Dp}) can
be written in terms of Eulerian perturbations;
\begin{equation}
\frac{ \delta \rho }{\rho} = \frac{1}{\Gamma_{\rm f}} \frac{\delta P}{P}
                           - \mtb{\xi} \cdot \mtb{A} \, . \label{eq:drho}
\end{equation}
The vector field $\mtb{A}$ is defined as
\begin{equation}
\mtb{A} \equiv \nabla \ln \rho - \frac{1}{\Gamma_{\rm f}} \nabla \ln P
 = \left( \frac{1}{\Gamma_{\beta}} - \frac{1}{\Gamma_{\rm f}} \right) \nabla \ln P \,
. \label{eq:Adef}
\end{equation}
Its magnitude is usually referred to as the Schwarzschild
discriminant, with an overall sign depending upon whether ${\bf A}$
points outward (positive discriminant) or inward (negative
discriminant).  Note that the background relation ~(\ref{eq:Gbdef})
has been used in the last equality of equation~(\ref{eq:Adef}).

For barotropic stellar oscillations a single equation of state
describes both the background and the perturbed matter. The adiabatic
indices $\Gamma_{\beta}$ and $\Gamma_{\rm f}$ are then equal and the
Schwarzshild discriminant vanishes, $\mtb{A}=0$. In this case, the
stellar structure does not sustain buoyancy restored modes so the
composition g-modes are absent from the
spectrum~\citep{1992ApJ...395..240R}.  In the nonbarotropic case,
$\mtb{A} \neq 0 $, the sign of the Schwarzschild discriminant
determines the convective stability of the motion.  Stable (unstable)
g modes are present in stars with negative (positive) Schwarzschild
discriminant~\citep{Tassoul:1978, 1989nos..book.....U}.

In our numerical evolutions, we do not use the Lagrangian displacement
$\mtb{\xi}$ as a dynamical variable. Instead we evolve the perturbed
proton fraction $\delta x_\p$.  The relevant dynamical equation
follows if we rewrite equation~(\ref{eq:drho}) using equation~(\ref{eq:Dxp});
\begin{equation}
\frac{ \delta \rho }{\rho} = \frac{1}{\Gamma_{\rm f}} \frac{\delta P}{P} -
\left( \frac{\Gamma_{\beta}}{\Gamma_{\rm f}} - 1 \right) \frac{\delta
  \chi_\p}{\rho} \, , \label{eq:drhochi}
\end{equation}
where we have defined
\begin{equation}
\delta  \chi_\p \equiv \delta x_\p / \left( \frac{d x_\p}{d \rho} \right) \, .
\end{equation}
We have also assumed that $x_\p = x_\p \left( \rho \right)$.

The composition gradient can be directly related to the deviation of
$\Gamma_{\rm f}$ from the background $\Gamma_{\beta}$. From
equation~(\ref{eq:dPEoS}) the pressure perturbation can be expressed as
\begin{equation}
\frac{\delta P}{P} = \Gamma_{\rm f} \frac{\delta\rho}{\rho}
                   + \Gamma_{\p} \frac{\delta x_\p}{x_\p} \, , \label{eq:dPfrac}
\end{equation}
where
\begin{equation}
\Gamma_{\p} \equiv \left. \frac{d\log P}{d \log x_\p} \right|_{\rho} \, .
\end{equation}
Comparing equations~(\ref{eq:drhochi}) and~(\ref{eq:dPfrac}) we obtain
the following relation between the proton fraction and the adiabatic
indices:
\begin{equation}
\Gamma_{\beta} -  \Gamma_{\rm f} = \Gamma_\p \frac{d \log x_\p}{ d \log \rho} \, .
\end{equation}

It is now straightforward to extend the models considered
by~\citet{2002MNRAS.334..933J} to account for composition variations.
Non-axisymmetric perturbations of rapidly, and uniformly, rotating
stratified neutron stars can be described by a system of five
equations.  This system is formed by the perturbed Euler, mass
conservation and frozen composition equations.  In the frame of the
rotating background and in terms of the dynamical variables $\delta
P$, $\delta\chi_\p$ and the flux $\mtb{f}=\rho \, \mtb{v}$, the
equations can be written
\begin{eqnarray}
\partial_t \mtb{f} & = & - \nabla \delta P - 2 \left(\mtb{\Omega} \times \mtb{f} \right)
                     + \frac{\nabla \ln P}{\Gamma_{\rm f}} \,  \delta P
                     + \left(1-\frac{\Gamma_\beta}{\Gamma_{\rm f}}\right)
                       \frac{\nabla P}{\rho} \delta \chi_\p \, ,                       \label{eq:dfdt} \\
\partial_t\delta P & = & - \Gamma_{\rm f}\frac{P}{\rho} \, \nabla \cdot  \mtb{f}
                     + \frac{1}{\rho} \left( \frac{\Gamma_{\rm f}}{\Gamma_\beta} - 1 \right)
                     \mtb{f} \cdot \nabla P \, ,      \label{eq:dPdt} \\
\partial_t \delta \chi_\p & = & - \frac{\mtb{f} \cdot \nabla P }{\Gamma_\beta P}  \, . \label{eq:dxpdt}
\end{eqnarray}
In writing down equation (\ref{eq:dfdt}) we have made the Cowling
approximation, i.e. neglected the forces due to the perturbed
gravitational potential.  Equation~(\ref{eq:dxpdt}) describes the
conservation of the proton fraction in a fluid element moving with the
fluid. It is equivalent to the time derivative of
equation~(\ref{eq:Dxp}).  For $\Gamma_{\beta} = \Gamma_{\rm f}$, the
system of equations~(\ref{eq:dfdt})-(\ref{eq:dxpdt}) reduces to the
barotropic perturbation equations considered in
\citet{2002MNRAS.334..933J}. In that case, $\delta \chi_\p$ is a
superfluous dynamical variable.

As is clear from the above equations, time evolution of the
non-axisymmetric perturbation equations is a three-dimensional problem
in space. However, by exploiting the axial symmetry of the stellar
background and the fact that we only consider linear perturbations,
all perturbed variables can be expanded in terms of a set of basis
functions $\left( \cos m \phi \, ,\sin m \phi \right)$, where $m$ is
the azimuthal harmonic index~\citep{1980MNRAS.190...43P}.  For
instance, in the case of the pressure we have
\begin{equation}
\delta P \left( t,r,\theta,\phi \right) = \sum_{m=0}^{m=\infty}
               \left[  \delta P_{m}^{+} \left( t,r,\theta\right)
                        \cos m \phi
                     + \delta P_{m}^{-} \left( t,r,\theta\right)
                        \sin m \phi
               \right] \, ,  \label{eq:dPexp}
\end{equation}
The perturbation equations then decouple with respect to $m$ and the problem
becomes two-dimensional. For any $m$, the non-axisymmetric oscillations
of a stratified neutron star requires the solution of a system of ten partial
differential equations for the ten variables $\left( \delta
P^{\pm}, \mtb{f}^{\pm}, \delta \chi_\p^{\pm} \right)$.

\section{Boundary Conditions} \label{sec:BC}

The evolution equations need to be complemented by boundary
conditions. At the stellar surface we require that the Lagrangian
perturbation of the pressure vanishes, i.e.,
\begin{equation}
\Delta P = \delta P + \mtb{\xi}  \cdot \nabla P = 0  \, . \label{eq:DP-bc}
\end{equation}
From the Euler equation for the stationary background
\begin{equation}
\nabla P = - \rho \mtb{\Omega} \times \left( \mtb{\Omega} \times \mtb{r} \right)
           - \rho \nabla \Phi   \, ,  \label{eq:Euler-bck}
\end{equation}
the condition~(\ref{eq:DP-bc}) reduces to
\begin{equation}
\Delta P = \delta P - \rho \mtb{\xi} \cdot \left( \mtb{\Omega} \times \mtb{\Omega} \times \mtb{r} 
           + \nabla \Phi \right) = 0 \, . \label{eq:Euler-bckB}
\end{equation}
In a polytropic star, the mass density vanishes on the stellar
surface. Therefore, if the Lagrangian displacement~$\mtb{\xi}$ assumes
finite values at the surface, according to
equation~(\ref{eq:Euler-bckB}) the pressure perturbation $\delta P$
behaves as $\rho$ near the surface.
In order to avoid divergent solutions in the numerical simulations, we
impose a zero surface condition for the mass flux perturbation
\begin{equation}
\mtb{f} = \rho \mtb{v} = 0 \, . \label{eq:scf}
\end{equation}
We consider initial conditions where the Lagrangian displacement is
finite on the stellar surface. Therefore, equation~(\ref{eq:scf})
implies that along the numerical evolution also the quantity $\rho
\mtb{\xi}$ remains zero on the surface.  As a result, from
equation~(\ref{eq:Euler-bckB}) the Eulerian pressure perturbation
satisfies a zero surface condition 
\begin{equation} 
\delta P = 0 \, .
\end{equation}
This condition ensures that the perturbed quantities remain 
regular at the surface. It is also straightforward to implement in the code.

We focus on the non-axisymmetric oscillations for $m \geq 2$. At the
origin, $r=0$, the regular behaviour of
equations~(\ref{eq:dfdt})-(\ref{eq:dxpdt}) is then guaranteed by the
conditions:
\begin{equation}
\delta P = \delta \chi_\p = \mtb{f} = 0 \, .
\end{equation}

\section{Oscillation modes of rotating stars}

Before we discuss the results of our evolutions, it is useful to
digress on the expected nature of the pulsation modes of a spinning
star.

Perturbations of spherical (non-spinning) stars can be decomposed in
two separate classes identified by their behaviour under the parity
transformation. Polar perturbations, which are expressed in terms of
the scalar spherical harmonics $Y_{l m}$ and their gradients $\nabla
Y_{l m}$, transform as $\left( -1 \right)^{l}$ under parity
inversion. Meanwhile, axial perturbations, which behave as $r \times
\nabla Y_{l m}$, change sign as $\left( -1 \right)^{l+1}$ under
parity. In a spherical star, the different multipoles do not couple.
An oscillation mode can be identified by the harmonic indices $l$ and
$m$ (together with some statement about the radial eigenfunction).

For rotating stars, the problem is more complicated.  First of all,
rotation removes the degeneracy in $m$ generating a richer spectrum
than in a non-rotating star. Secondly, the identification of a mode
with a single $l$ is well defined only for non-rotating
models. Nevertheless, each individual mode can be tracked by
increasing the rotation rate step by step. It is also important to
note that, even though the spectral properties of an oscillation mode
change continuously along a sequence of rotating models, its overall
parity is conserved.

In this paper we  discuss the three main classes of oscillation modes.
They are classified according to the
dominant restoring force that acts on the perturbed fluid
elements~\citep{Cowling:1941co}.  For polar modes the high
frequency part of the spectrum, roughly above $1$~kHz, is occupied by
the fundamental f-mode and the acoustic p-modes.  In stars where
composition or thermal gradients are present,  low-frequency g-modes
are also present. These are restored by gravity through the buoyancy
force. These various modes are non-trivial already in a non-rotating star.
A polar mode labeled by the harmonic index $l$ in
a non-rotating configuration is  split by rotation into $2l + 1$
distinct modes (identified by the value of $m$).

In contrast, there are no non-trivial axial modes in a non-rotating
fluid star.  The axial modes form a zero frequency subspace that
describe stationary convective currents~\citep{1999ApJ...521..764L}.
In a rotating configuration, these modes become non-trivial thanks to
the Coriolis force.  Collectively, they are referred to as inertial
modes. A general mode in this rotation-dominated class can have a
mixture of axial and polar components. However, since each mode has a
distinct parity one can distinguish two separate classes of inertial
modes. Following \citet{1999ApJ...521..764L} we refer to these sets as
polar- and axial-led. In the family of axial-led inertial modes, you
find the so-called r-modes. They are particularly simple in that they
are purely axial in the slow-rotation limit. For barotropic stars only
the $l=m$ r-modes exist, while in non-barotropic stars $l\neq m$
r-modes also appear. An interesting question concerns the fate of
these r-modes in the barotropic limit. Some aspects of this question
were resolved by \citet{1999ApJ...521..764L}, but interesting
questions remain. Consider, for example, a weakly stratified star. In
a fast spinning configuration the buoyancy should be dominated by the
Coriolis force and the low frequency modes should essentially have an
inertial nature. For lower rotation rates there should be a transition
to a buoyancy dominated regime, where the g-modes become distinct.  Of
course, the relative strength of the Coriolis force and the buoyancy
will vary throughout a star. As we will discuss in
Section~\ref{sec:Resnobar} one would expect to find distinct regions
where buoyancy dominates the Coriolis force, and other regions where
it is less important~\citep{2000A&A...354...86D}.  It is then
important to understand these two regions better.  One would certainly
want to know if it is possible to track individual modes from one
regime to the other. Ideally, this would shed some light on the
relationship between the inertial modes of a barotropic model and the
g-modes of a non-barotropic system.

Let us finally make a connection between the various classes of modes
and our evolution problem. In the evolutions, we do not distinguish
between axial and polar perturbations. Instead, we use the fact that
perturbations of an axisymmetric background can be decomposed into two
parity classes that satisfy different conditions in the equatorial
plane. For the first class, which we will refer to as type I parity,
$\delta P^+ , \delta \chi_\p^+ , f_r^+, f_{\phi }^+$ are all even
under reflection with respect to the equatorial plane, while
$f_{\theta}^- $ is odd. The $l=m$ fundamental modes and the polar-led
inertial modes belong to this class.  Conversely, for the second
class, that we will refer to as type II parity, $\delta P^- , \delta
\chi_\p^- , f_r^-, f_{\phi}^-$ are odd and $f_{\theta}^+ $ is
even. Typical modes of a rotating star that belong to this parity
class are the $l+1=m$ fundamental modes and the axial-led inertial
modes.

\begin{figure}
\begin{center}
\includegraphics[height=90mm]{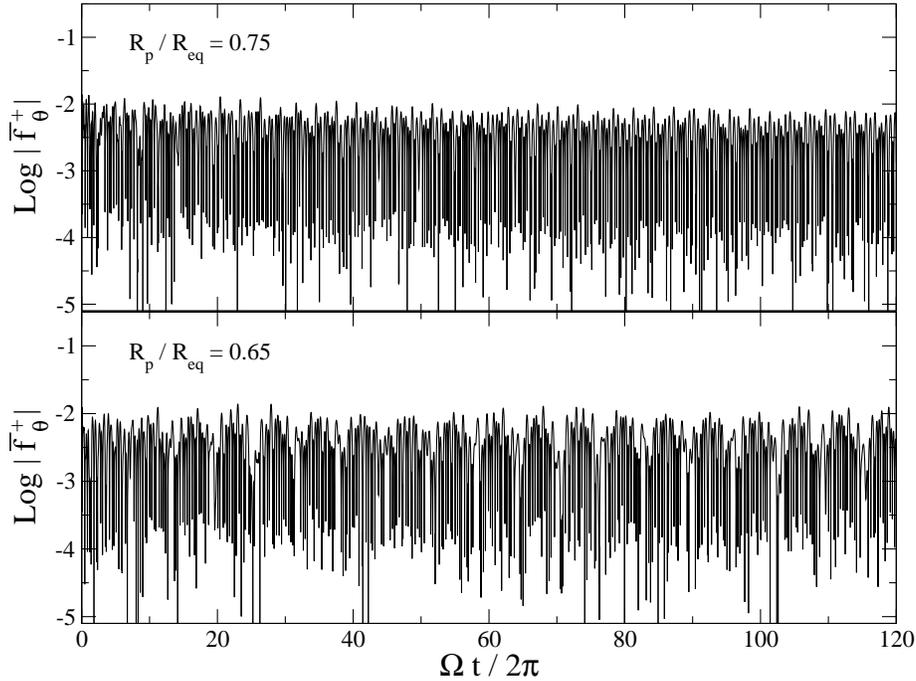} 
\caption{This figure shows the time evolution of the integrated flux
component $\bar{f}_{\theta}^{+}$, defined in
equation~(\ref{eq:intfth}), for two fast rotating models with $R_{p} /
R_{eq} = 0.75$ and $R_{p} / R_{eq} = 0.65$. Time is given in
dimensionless units as a number of the star's rotational
periods. These two simulations show the stability and the small
numerical dissipation of our numerical code. 
\label{fig:code-test}}
\end{center}
\end{figure}

\section{The code \label{app:numcode}}

After decomposition with respect to $\phi$, the dynamical problem has
two spatial dimensions, represented by the angle $\theta$ and the new
radial coordinate $x=x\left( r, \theta \right)$. This radial
coordinate is fitted to surfaces of constant pressure of the
background star and it is normalised to unity on the equatorial
radius.  The choice of coordinate simplifies the implementation of the
surface boundary condition in a perturbed star
considerably~\citep{2002MNRAS.334..933J}.  The spatial variables $(x,
\theta)$ are then discretized on an evenly spaced two-dimensional
grid, where $0 \leq x \leq 1$ and $ 0 \leq \theta \leq \pi / 2 $.

We evolve equations~(\ref{eq:dfdt})-(\ref{eq:dxpdt}) in time with a
Mac-Cormack algorithm, a second order accurate scheme that involves a
predictor and a corrector step. In order to prevent spurious numerical
oscillations we use a fourth-order Kreiss-Oliger dissipation with a
parameter that depends on the numerical grid size.  The perturbation
equations are evolved for the internal points of the numerical grid,
while boundary conditions are used to update the perturbations at the
origin, the rotation axis, the surface and the equator.  In order to
use the fourth-order Kreiss-Oliger dissipation also near the grid
boundary one ghost point is added to each grid slice.  Numerical
simulations are then carried out on $(x, \theta)$ grids with $60\times
32$ points. Test simulations on finer grids with $60\times 64$ or $120
\times 32$ points do not give significant improvements on the results.
The numerical evolutions are stable for several hundred rotation
periods. We extract the required mode frequencies from the evolutions
by performing a Fast Fourier Transformation (FFT) of the time series
of the perturbations.

In order to have global information of the pulsations of the star, the
perturbation variables can be integrated on the spatial grid (using a
two dimensional integral). For instance, the integrated flux
perturbation component $f_{\theta}^{+}$ is given by
\begin{equation}
\bar{f}_{\theta}^{+} \equiv \int_{0}^{\pi/2}
d\theta \int_{0}^{R(\theta)} f_{\theta}^{+} r dr \, , \label{eq:intfth}
\end{equation}
where $R(\theta)$ is the radius of the star at any angle $\theta$.
This integrated quantity is shown in Fig~\ref{fig:code-test}, for two
fast rotating stellar models with axis ratio 0.75 and 0.65, for
evolutions lasting up to 120 rotation periods.  This figure
illustrates the codes stability as well as the very small numerical
dissipation. Similar characteristics are observed for other
perturbation quantities.  These results demonstrate a significant
improvement on the results reported in~\citet{2002MNRAS.334..933J}.
This improvement in stability allows us to analyse correspondingly
longer stretches of data, providing more accurate localisation of
modes in frequency space.

\begin{figure}
\begin{center}
\includegraphics[width=90mm]{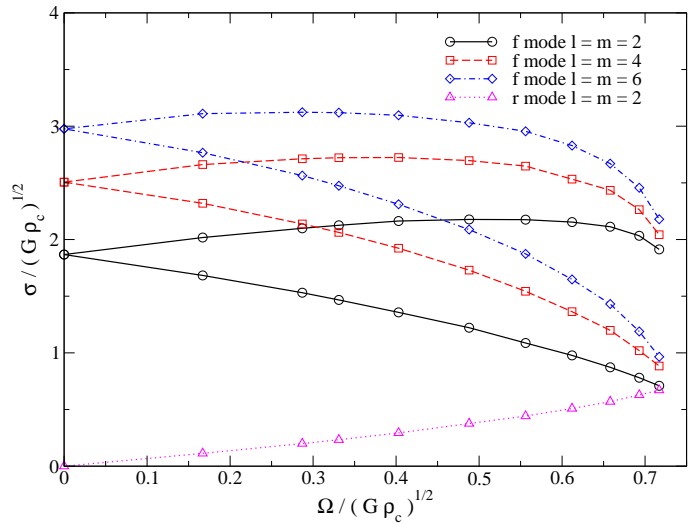}
\caption{This figure displays the frequencies of the $l=m$ f- and
  r-modes as a function of the rotating rate for the
  $\Gamma_{\beta}=2$ barotropic sequence of rotating stars. The mode
  frequencies are determined in the rotating frame.  All frequencies
  are given in units of $\sqrt{G \rho_c}$, where $\rho_c$ is the mass
  density at the stellar centre.  These results test our numerical
  code in the limit of barotropic stars. The mode frequencies are in
  good agreement with those reported by \citet{2002MNRAS.334..933J}.
  \label{fig:baro-f-r-mode}}
\end{center}
\end{figure}

\section{Results} \label{sec:Results}

\subsection{Initial Data} \label{sec:ID}

Once we fix the azimuthal harmonic index $m$ the code evolves linear
oscillations on an axisymmetric background. This means that modes with
any value of $l$ can be excited. This makes the problem different from
a perturbation analysis in the frequency domain, where one tends to
have some control over the specific angular dependence of the modes
that are studied.  The spectral content of the evolutions, which is
extracted from an FFT of the time evolved perturbation variables,
depends on the initial data used to excite the motion.

On the initial time slice, a single oscillation mode can, in
principle, be excited by providing the correct eigenfunctions for the
perturbation variables. However, this is not a practical strategy
since it requires the mode-problem to be solved already. An
alternative is to use an eigenfunction recycling
method~\citep{Stergioulas:2003ep, Dimmelmeier:2005zk}.  In this study
we do not need these refinements.  We are interested in a multi-mode
analysis, so we can simply choose arbitrary initial perturbations that
excite as many modes as possible in a single numerical simulation.

We excite type I parity perturbations with
a Gaussian radial profile in the pressure perturbation:
\begin{equation}
\delta P = \rho \, \exp{\left[ - \left( \frac{r-r_0 }{ q  R\left( \theta \right)} \right) ^2 \right] }
           Y_{l l} \left(\theta,\phi\right) \, , \label{eq:fID}
\end{equation}
where $R\left( \theta \right)$ is the stellar radius at polar angle
$\theta$, and the parameters $r_0$ and $q$ determine the centre of the
Gaussian and its width, respectively. The function $Y_{l l}
\left(\theta,\phi\right)$ is the $l=m$ spherical harmonic which
describes a typical polar mode angular pattern of a spherical star.
The initial conditions of the mass flux $\mtb{f}$ and proton fraction
$\chi_\p$ perturbations are set to zero.  Type II parity perturbations
are instead excited by the flux perturbation:
\begin{equation}
\mtb{f} = \rho \, \exp{\left[- \left( \frac{r-r_0 }{ q  R\left( \theta \right)} \right) ^2 \right] }
           Y_{l l}^B \left(\theta,\phi\right) \, ,
\end{equation}
where $Y_{l l}^B \left(\theta,\phi\right)$ is a magnetic spherical
harmonic~\citep{1980RvMP...52..299T}. In this case, the pressure and
the proton fraction perturbations both vanish on the initial time
slice.  Note that for both type I and type II parity, the coupling of
the perturbed quantities in equations
(\ref{eq:dfdt})--(\ref{eq:dxpdt}) ensures these these simple choices
of initial data sets do in fact excite non-trivial perturbations in
all quantities.

\subsection{The Barotropic Case} \label{sec:Resbar}

We first consider the problem for barotropic stars. This serves as a
useful test-case for the reliability of our evolutions. The results
also illustrate the improvements that we have made on the previous
analysis \citep{2002MNRAS.334..933J}.

In barotropic rotating stars the oscillation spectrum is formed by
pressure, fundamental and inertial modes. Rotation removes the
degeneracy (in $m$) of the non-axisymmetric pressure and f-modes and
generates the new family of inertial modes, which are restored by the
Coriolis force. The main properties of the inertial modes are that
they vanish linearly in the limit of zero stellar rotation and that
they have mixed nature. Even in the slow rotation limit the velocity
perturbations of the inertial modes have both axial and polar
parts~\citep{1999ApJ...521..764L}. However, one can still identify two
independent classes of inertial modes. The axial-led class with parity
$(-1)^{m+1}$ and the polar-led modes with parity $(-1)^{m}$. The
velocity perturbation of an axial-led inertial mode can couple only
with axial terms having $l=m,m+2,\dots$ and with polar terms having
$l=m+1,m+3,\dots$. In the case of a polar-led inertial mode, the
velocity perturbation couples only with $l=m,m+2,\dots$ polar terms
and $l=m+1,m+3,\dots$ axial terms.  A particular subset of the
axial-led inertial modes is formed by the r-modes, which have purely
axial velocity perturbations in the slow-rotation limit. For
barotropic stars only $l=m$ r-modes can exist.

\begin{table}
\begin{center}
\caption{\label{tab:baro-inertial} Inertial mode frequencies for
barotropic rotating models with $\Gamma_{\beta}=2$ . The purely axial
$l =m=2$ r-mode is denoted by the symbol ${}^{l} \rm{r}$ and the
inertial modes by ${}^{l} \rm{i} _{k}$, where $k$ labels different
modes with the same $l$.  The second column shows the overall mode
parity, i.e. whether the mode is axial or polar-led.  Frequencies
determined by~\citet{1999ApJ...521..764L} are labeled with the acronym
L-F and listed in the third column, while those extracted from our
numerical code are shown in the last two columns for two slowly
rotating models.  In the formalism used by L-F, positive and negative
eigenfrequencies correspond to counter and co-rotating modes
respectively.  In an FFT of a time evolved perturbation variable, they
all appear positive, but their counter- or co-rotating character can
easily be established by comparison with the L-F results and the
dependence on the rotation of the star.}
\begin{tabular}{c c c c  c }
\hline
   Mode      & Parity  & $\sigma  / \Omega $ & $\sigma / \Omega$ &  $\sigma  / \Omega$   \\
             &         &   L-F   &  $ R_p/R_{eq} = 0.983 $  &  $ R_p/R_{eq} = 0.95 $ \\
\hline
  ${}^2\rm{r}$    &  a      &  0.667  &  0.676  &  0.695  \\
  ${}^3\rm{i}_1$  &  p      & -0.557  &  0.562  &  0.549  \\
  ${}^3\rm{i}_2$  &  p      &  1.100  &  1.101  &  1.116  \\
  ${}^4\rm{i}_1$  &  a      & -1.026  &  1.021  &  1.036  \\
  ${}^4\rm{i}_2$  &  a      &  0.517  &  0.508  &  0.536  \\
  ${}^4\rm{i}_3$  &  a      &  1.358  &  1.363  &  1.375  \\
  ${}^5\rm{i}_1$  &  p      & -1.273  &  1.291  &  1.296  \\
  ${}^5\rm{i}_2$  &  p      & -0.275  &  0.252  &  0.258  \\
  ${}^5\rm{i}_3$  &  p      &  0.863  &  0.834  &  0.884  \\
  ${}^5\rm{i}_4$  &  p      &  1.519  &  1.512  &  1.519  \\
  ${}^6\rm{i}_1$  &  a      &  0.422  &  0.391  &  0.409  \\
\hline
\end{tabular}
\end{center}
\end{table}


\begin{figure}
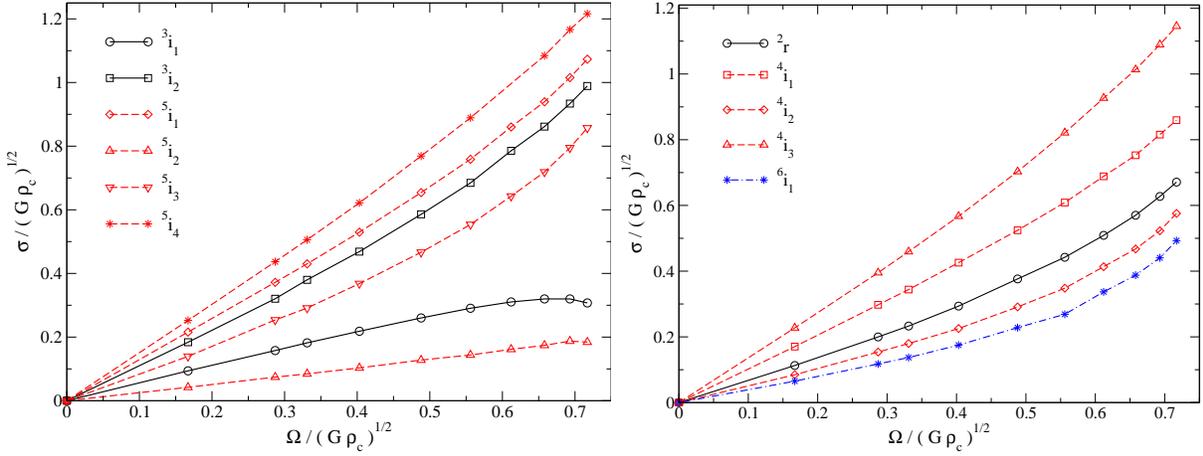

\begin{center}
\includegraphics[height=60mm]{fig3a.eps}
\includegraphics[height=60mm]{fig3b.eps}
\caption{Rotating frame frequencies of polar-led inertial modes (left
panel) and the r-mode and other axial-led inertial modes (right panel)
as a function of the rotation rate for $\Gamma_{\beta}=2$ barotropic
models. The inertial mode frequencies shown in this figure reproduce
with good accuracy the results presented
by~\citet{2002MNRAS.334..933J}. In the limit of slow rotation, the
mode frequencies are also in good agreement with the results reported 
by~\citet{1999ApJ...521..764L}, see Table~\ref{tab:baro-inertial}.
\label{fig:baro-in-polar-led}}
\end{center}
\end{figure}

The non-axisymmetric modes of barotropic Newtonian rotating stars have
been extensively investigated in literature. We tested our numerical
code by reproducing the results of the previous code developed
by~\citet{2002MNRAS.334..933J}, finding a very good agreement.  For
stars with $\Gamma_{\beta}=\Gamma_{\rm f}=2$, we show in
Fig.~\ref{fig:baro-f-r-mode} the rotational splitting of the $l=m$
f-modes and the $l=m=2$ r-mode. The final stellar model in the
sequence is rotating very fast, at about 99 percent of the Kepler
limit.  The new code performs significantly better that the one
discussed in~\citet{2002MNRAS.334..933J}.  In particular, the improved
stability allows us to study configurations that rotate closer to the
break-up limit. Being able to carry out longer simulations, we can
also improve on the precision of the extracted mode frequencies.

In addition to the f-mode, many inertial modes are excited during a
typical evolution. For type I parity perturbations and $m=2$, we can
identify polar-led $l=3$ and $l=5$ inertial modes. These are shown in
the left panel of Fig.~\ref{fig:baro-in-polar-led} for several
barotropic rotating stars. For $m=2$ type II parity initial
conditions, axial-led $l=4$ and $l=6$ inertial modes and the $l=m=2$
r-mode are excited and their frequencies are shown in the right panel
of Fig.~\ref{fig:baro-in-polar-led}.  In order to identify and test
the inertial mode frequencies, we have compared them to the
frequencies determined by~\citet{1999ApJ...521..764L}. As is clear
from the data in Table~\ref{tab:baro-inertial}, the agreement is
better than $6\%$ for slowly rotating barotropic models with axis
ratio $R_p /R_{eq} = 0.983$ and $R_p / R_{eq} = 0.95$.

It is worth noting that, some of the eigenfrequencies in
Table~\ref{tab:baro-inertial} are negative as these modes are
co-rotating with the stellar fluid. Meanwhile, the frequencies that we
extract will always be positive, as we are using an FFT of the time
evolved perturbations.  The co- and counter-rotating character of the
various modes can be established by comparison with the results
determined by~\citet{1999ApJ...521..764L}.  In some cases, the nature
of the mode can also be inferred from the dependence on the rotation
rate of the star.  As can be seen from, for example,
Figure~\ref{fig:baro-in-polar-led} some of the co-rotating mode
frequencies (in the rotating frame) tend to decrease at fast rotation
rates. Meanwhile, no counter-rotating modes show this effect.

\subsection{Stratified Stars} \label{sec:Resnobar}
\begin{table}
\begin{center}
\caption{\label{tab:gmodes-YL} Comparison of the first two $l=2,3,4$
g-modes frequencies with the~\citet{2000ApJS..129..353Y} results. The
star is non-rotating and have adiabatic indices $\Gamma_{\beta}=2$ and
$\Gamma_{\rm f} = 2.0004$. The frequencies are given in units of
$\sigma / \sqrt{G \rho_c}$ and the Yoshida and Lee values are denoted
with YL. }
\begin{tabular}{c c c c c   }
\hline
  $  l  $  &  $ \rm{g}_{1}$  & $\rm{g}_{1}$  & $ \rm{g}_{2} $
& $ \rm{g}_{2} $   \\
                &  YL  &   &  YL  &    \\
\hline
   2            & 0.0189  &  0.0187  & 0.0129 & 0.0123   \\ 
   3            & 0.0227  &  0.0229  & 0.0162 & 0.0159   \\ 
   4            & 0.0255  &  0.0252  & 0.0189 & 0.0188   \\
\hline
\end{tabular}
\end{center}
\end{table}
The oscillation spectrum of a stratified star is, in a sense, richer
than that of a barotropic model. Fluid elements that have moved away
from the equilibrium position can produce composition gradients in the
stellar matter. For small displacements, buoyancy acts as restoring
force by producing an oscillatory motion of the perturbed fluid
element. These oscillation modes are called composition gravity modes
(g-modes). They must been discerned from thermal g-modes, which are
due to temperature gradients. The latter are only relevant in newly
born neutron stars, while the composition modes are relevant also in
mature systems.  Gravity modes have a polar nature, appear in the
low-frequency band of the oscillation spectrum, typically around 100
Hz, and exist even for non-rotating stellar models. Like f- and
p-modes, each g-mode can be labeled by the harmonic indices $l$ and
$m$ together with the number of nodes in the radial eigenfunction.  We
will label individual g-modes using the notation ${}^l\rm{g}_{\rm k}$,
where k labels different modes sharing the same value of the index
$l$.  Additionally, we will denote a retrograde (as opposed to
prograde) mode as ${}^l\rm{g}^r_{\rm k}$

There is an overlap in frequency space between the g-modes and the
inertial modes. In fact, since both sets of modes emerge from the
degenerate trivial modes of a non-rotating barotropic model, one might
expect them to be intimately related.  In order to understand this
relationship better, we will investigate the low frequency
oscillations of non-barotropic rotating neutron stars. The aim is to
clarify whether inertial and g-modes are both present in the spectrum
and to what extent it is possible to distinguish them. We also want to
understand their behaviour in fast rotating models.

\begin{table}
\begin{center}
\caption{\label{tab:gmodes} The frequencies of the first three $l=2$
g-modes for a non-rotating non-barotropic model with $\Gamma_{\beta}=2$ and
three different values of $\Gamma_{\rm f} = 2.05, 2.2,2.4$. The frequencies are
given in units of $\sigma / \sqrt{G \rho_c}$. This comparison of modes
determined with a frequency (FD) and time domain (TD) approach provide a useful
reliability check of our time evolutions.}
\begin{tabular}{c c c c c c c  }
\hline
  $ \Gamma_{\rm f} $  &  $ {}^2\rm{g}_1$  & ${}^2\rm{g}_{1}$  & $ {}^2\rm{g}_{2} $
& $ {}^2\rm{g}_{2} $  & $ {}^2\rm{g}_{3} $ & ${}^2\rm{g}_{3}$  \\
                &  FD & TD & FD & TD &  FD & TD  \\
\hline
   2.05         &  0.195  & 0.208  & 0.134  & 0.141  &  0.102  & 0.111  \\
   2.2          &  0.383  & 0.396  & 0.266  & 0.271  &  0.205  & 0.209  \\
   2.4          &  0.531  & 0.530  & 0.370  & 0.367  &  0.287  & 0.284  \\
\hline
\end{tabular}
\end{center}
\end{table}

\begin{figure}
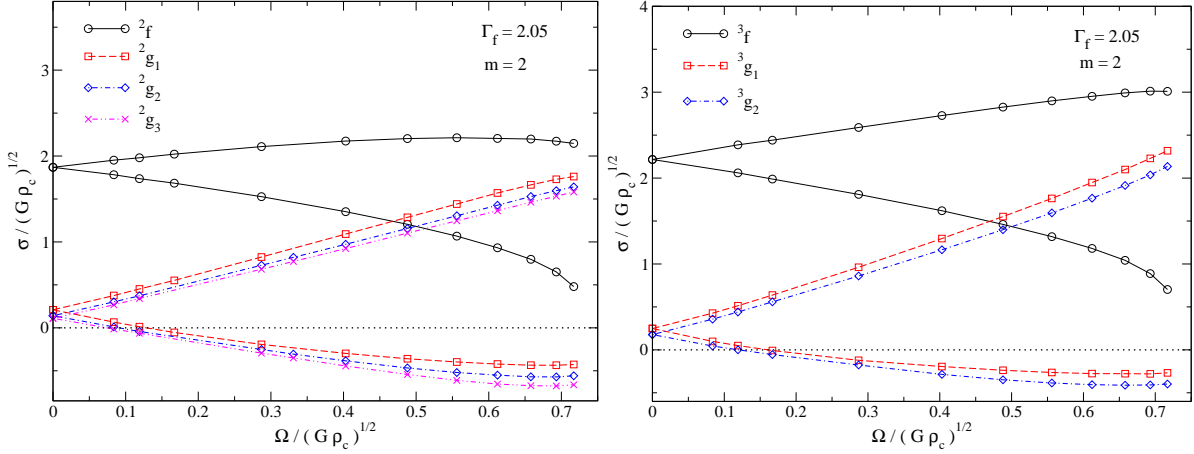

\begin{center}
\includegraphics[height=60mm]{fig4a.eps}
\includegraphics[height=60mm]{fig4b.eps}
\caption{Results for non-barotropic rotating stellar models with
$\Gamma_{\beta}=2$ and $\Gamma_{\rm f}=2.05$. The left panel shows the
frequencies of the $l=m=2$ f- and g-modes, while the right panel shows
the corresponding modes for $l=3$ and $m=2$.  In order to emphasize
the rotational splitting of the g-modes and the onset of the CFS
instability, the frequencies are shown in the reference frame of an
inertial observer.
\label{fig:Gf2.05-f-gmodes-polar}}
\end{center}
\end{figure}

These issues can be directly related to the dynamics of a perturbed
fluid element and the balance between the two restoring forces that
are acting on it, the Coriolis and the buoyancy force. In slowly
rotating and stratified neutron stars with large composition
gradients, buoyancy is expected to dominate the Coriolis force. As a
result, a typical mode should appear as a g-mode.  For higher rotation
rates, the effect of the Coriolis force increases and may dominate the
buoyancy force beyond a given stellar spin. In this regime, the
oscillation mode should behave as an inertial mode.  The threshold
between these two regimes depends on the magnitude of the composition
gradient and must obviously be smaller than the mass shedding limit of
the star to be of any interest.

The order of magnitude of the composition g-mode frequencies can be
estimated by the Brunt-V\"{a}is\"{a}l\"{a} frequency $N$, which is
defined by~\citep[see e.g.,][]{ 1989nos..book.....U}
\begin{equation}
N^2 \equiv \frac{\nabla P}{\rho} \cdot \mtb{A} = \left(
\frac{1}{\Gamma_{\beta}} - \frac{1}{\Gamma_{\rm f}} \right) \frac{
\left| \nabla P \right| ^2}{\rho P } \, . \label{eq:Nfreq}
\end{equation}
For non-rotating stars this frequency can be written 
\begin{equation} N^2 = - g A = - g \left( 1 -
\frac{\Gamma_{\beta}}{\Gamma_{\rm f}} \right) \frac{d \ln \rho}{d r} \,
,  \label{eq:Nfreq-nr}
\end{equation}
where the equilibrium configuration equations have been used to replace 
the pressure gradient with the gravitational acceleration.
An averaged value of equation~(\ref{eq:Nfreq-nr}) can be estimated by
approximating the gravitational acceleration as $g \simeq G M / R^2
\simeq \left(4 \pi / 3\right) G R \langle \rho \rangle $, where
$\langle \rho \rangle$ is the mass density average, and the density
gradients as $d \rho / dr \simeq - \rho_c / R$.  As a result, the
averaged and dimensionless Brunt-V\"{a}is\"{a}l\"{a} frequency $\hat N
\equiv N / \sqrt{G \rho_c}$ is given by:
\begin{equation}
\langle \hat{N} \rangle  \, \simeq \sqrt{ \frac{4\pi}{3} } \left( 1 -
                          \frac{\Gamma_{\beta}}{\Gamma_{\rm f}} \right)
                          ^{1/2} \, .
\label{NN}\end{equation}
We have tested equation~(\ref{NN}) with the average of
equation~(\ref{eq:Nfreq}) for the non-rotating and non-barotropic
models used in this paper.  The results agree to better than $5\%$.

\begin{figure}
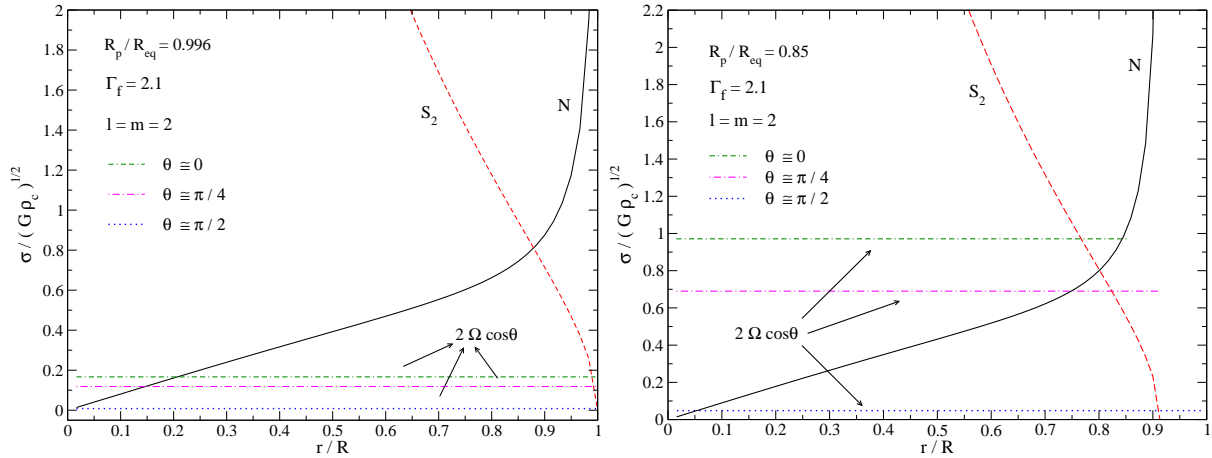

\begin{center}
\includegraphics[height=60mm]{fig5a.eps}
\includegraphics[height=60mm]{fig5b.eps}
\caption{This figure displays the propagation diagram of two
stratified models with $\Gamma_\f = 2.1$ and $\Omega / \sqrt{G \rho_c}
= 0.084$ in the left panel and $\Omega / \sqrt{G \rho_c} = 0.488$ in
the right panel. The Brunt-V\"{a}is\"{a}l\"{a} frequency $N$, the Lamb
frequency $S_l$ and the Coriolis term $2\Omega\cos\theta$ of the
dispersion relation~(\ref{eq:disprel}) are given in $\sqrt{G \rho_c}$
units. On the horizontal-axis the radial coordinate $r$ is normalized
with the stellar radius $R$. For simplicity, the $N$ and $S_l$
frequencies are presented in the two panels for $\theta=\pi/4$, while
three directions are shown for the Coriolis term, respectively $\theta
= 0, \pi/4$ and $\pi/2$. The two propagation diagrams illustrate the
regions of the star where the various restoring forces
dominate. Increasing the stellar spin, the Coriolis term becomes
stronger than the Brunt-V\"{a}is\"{a}l\"{a} frequency in almost all
the volume of the star. Only in a small region near the equator,
buoyancy can dominate even in rapidly rotating stars.
\label{fig:prop-diag}}
\end{center}
\end{figure}

We want to study the effects of composition gradients on the spectrum
in detail.  We consider non-barotropic models with $\Gamma_{\beta} =
2$ and frozen composition indices between $\Gamma_{\rm f} = 2.05$ and
$\Gamma_{\rm f} = 2.4$ in order to investigate stars with low and high
composition gradients.
It is useful to begin by estimating at what rotation rate the Coriolis
force will start to dominate the buoyancy.  When $\Gamma_{\rm f} = 2.05$,
equation~(\ref{NN}) suggests that $N \sim 2 \Omega$ for a model with
$\Omega/\sqrt{G\rho_c}=0.16 $. For models rotating much faster
than this, we would expect the Coriolis force to dominate. As a
result, one would expect the low-frequency modes of such models to
have a predominantly inertial character.  As we will see later, the
trend suggested by this rough estimate is brought out by the detailed
mode-calculations.

\begin{figure}
\begin{center}
\includegraphics[height=60mm]{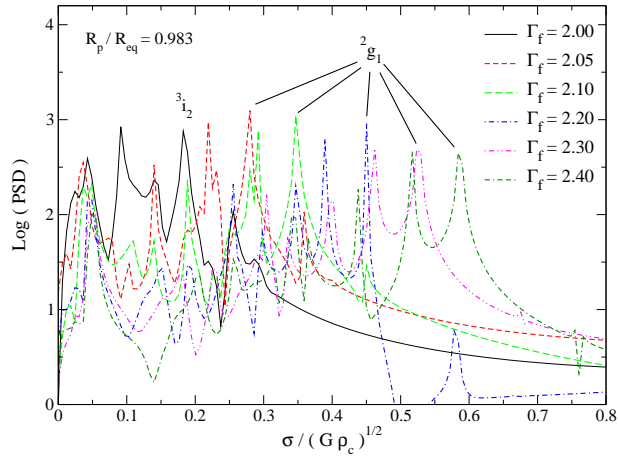}
\caption{ This figure shows the Power Spectral Density (PSD) of the
integrated flux perturbation component $f_{\theta}^{+}$ for a set of
slowly rotating barotropic and non-barotropic models. All these models
have axis ratio $R_{p} / R_{eq} = 0.983$ (corresponding to $\Omega /
\sqrt{G\rho_c} = 0.167$), but different frozen composition adiabatic
index $\Gamma_{\f}$.  The solid line corresponds to a barotropic star
with $\Gamma_{\f}= 2$. The counter-rotating ${}^2\rm{g}_1$ mode has
been determined, from left to right in the figure, for stars with
$\Gamma_{\f} = 2.05, 2.1,2.2,2.3,2.4$. This figure shows the
dependence of the ${}^2\rm{g}_1$ mode frequency on the composition
gradients. In the limit of $\Gamma_\f \rightarrow \Gamma_{\beta}$, the
counter-rotating ${}^2\rm{g}_1$ mode seems to approach the
${}^3\rm{i}_2$ inertial mode. This behaviour is clearer in
Fig~(\ref{fig:g1r-i31}).
\label{fig:Sp}}
\end{center}
\end{figure}
\begin{figure}
\begin{center}
\includegraphics[height=60mm]{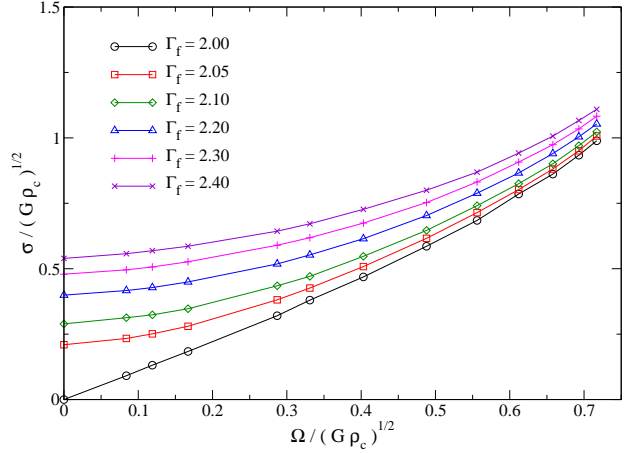}
\caption{ This figure displays the variation of the counter rotating
  ${}^2\rm{g}_{1}^{r}$ mode with the adiabatic index $\Gamma_{\f}$ for
  a sequence of rotating stellar models.  The polar-led inertial mode
  ${}^3\rm{i}_{2}$ frequencies of the barotropic models with
  $\Gamma_{\f}=2$ are shown as a circle-solid line.  This figure
  suggests that in rapidly rotating stars the counter rotating
  ${}^2\rm{g}_{1}^{r}$ mode behaves similarly to the barotropic
  ${}^3\rm{i}_{2}$ inertial mode. This is expected when the rotation
  rate is larger than the Brunt-V\"{a}is\"{a}l\"{a} frequency $N$,
  namely $2 \Omega > N$, and the Coriolis force starts to dominate
  over the buoyancy force.
\label{fig:g1r-i31}}
\end{center}
\end{figure}

The non-barotropic evolution code has been tested against two
eigenvalue codes developed by~\citet{2000ApJS..129..353Y}
and~\citet*{2008PhRvD..77b4029P}, respectively.
\citet{2000ApJS..129..353Y} solve for linear perturbations of slowly
rotating neutron stars in Newtonian gravity, whereas
~\citet{2008PhRvD..77b4029P} study the pulsation spectra of slowly
rotating relativistic stars in the Cowling approximation.  We have
compared the first two g-modes frequencies with the non-rotating and
weakly stratified Newtonian model of~\citet{2000ApJS..129..353Y},
which has $\Gamma_{\beta}=2$ and $\Gamma_{\f}= 2.0004$. The relevant
frequencies, which are given in Table~\ref{tab:gmodes-YL}, show a very
good agreement even though we are using the Cowling approximation.
The second $l=2$ g-mode have an error of four percent, while the
others are accurate to within two percent. This difference is due the
low frequency of the second $l=2$ g-mode that requires a longer time
evolution for extracting a more accurate value with an FFT.

We have further tested our code on neutron star models with larger
 composition gradients by using the~\citet{2008PhRvD..77b4029P} code.
 Since the time evolution code used in this paper is Newtonian, we
 have made a comparison between the two codes for a spherically
 symmetric background model with low compactness. For a polytropic
 model with $\Gamma_{\beta} = 2$ and a star with central density
 $\rho_c = 4.5 \times 10^{14}~\textrm{g cm}^{-3}$, the
 Tolman-Oppenheimer-Volkoff equations determine a non-rotating star
 with mass $M=1.17 M_{\odot}$ and radius $R=16$~km, giving a
 compactness ratio of $M/R = 0.11$.  In terms of the dimensionless
 quantity used in our time domain code, this model has $M/\left(
 \rho_c R^3 \right) = 1.262$ which agrees very well (within one
 percent) with the non-rotating model in Table~\ref{tab:back-models}.
 By using these two approaches, we determined the g-mode frequencies
 of non-rotating stratified neutron stars in
 Table~\ref{tab:gmodes}. For non-barotropic stars with $\Gamma_{\rm
 f}=2.05$ the frequencies of the first three modes agree to better
 than eight percent, while for $\Gamma_{\rm f}=2.4$ they agree to
 better than one percent.  This level of agreement is satisfactory
 given that the calculations are carried out within completely
 different frameworks. Thus we are confident that our time-evolutions
 are able to provide accurate g-mode results.

The rotational splitting of the $l=m=2$ f- and g-modes for stratified
stars with $\Gamma_{\f} = 2.05$ is shown, in the frame of an inertial
observer, in Fig.~\ref{fig:Gf2.05-f-gmodes-polar}. The left panel
represents modes with type I parity, while the right panel shows
results for type II parity. The two classes of modes clearly behave in
a similar fashion.  Results for other values of $\Gamma_{\rm f}$ are
similar. Generally, we observe that the frequencies of the f- and
g-modes tend to increase for stars with larger composition gradients,
i.e. larger $\Gamma_{\rm f}$. From the results in
Fig.~\ref{fig:Gf2.05-f-gmodes-polar} it is easy to read off when the
low-order g-modes become susceptible to the gravitational-radiation
driven CFS instability~\citep{1978ApJ...222..281F,Andersson:2002ch}.
The instability condition is that the mode-frequency changes sign
according to an inertial observer.  The results in the figure suggest
that this happens for $\Omega/\sqrt{G\rho_c} \approx 0.12$. This is in
accordance with the typical values expected for the g-mode CFS
instability.

In fact, we can compare our result to an order of magnitude estimate
based on the result of~\citet{1999MNRAS.307.1001L}. He estimates that
the rotation rate of the g-mode instability onset corresponds to
\begin{equation}
\Omega = 0.68 \, \sigma_0 \, , \label{eq:OmegaCFS}
\end{equation}
where $\sigma_0$ is the g-mode frequency for the non-rotating model.
From this we find that one would expect the instability of the
${}^2\rm{g}_1$ mode for the $\Gamma_{\rm f} = 2.05$ model to be
triggered in a star rotating at $\Omega/\sqrt{G\rho_c} \approx
0.14$. This is close to the (presumably more accurate) value
determined with our numerical code.

However, the g-mode instability is not thought to be very important
since the g-modes are not radiating gravitational waves efficiently.
Viscosity is expected to suppress the instability in more realistic
models. In comparison, the f-mode, which radiates gravitational waves
efficiently does not go unstable until the star spins extremely fast.
For stellar models with $\Gamma_{\beta}=2$, the $l=m=2$ f-mode is not
expected to become unstable before the mass shedding limit rotational
rate is reached~\citep{1990ApJ...355..226I}. As is clear from
Fig.~\ref{fig:Gf2.05-f-gmodes-polar}, the f-mode instability point
occurs for faster spins than we are able to study. This accords well
with previous results in the literature.

Let us now study the low frequency band of rotating non-barotropic stars. A
local analysis of uniformly rotating stars in the Cowling
approximation leads to the following dispersion relation for
low-frequency waves~\citep{1989nos..book.....U}:
\begin{equation}
\sigma ^2 \simeq \frac{ N^2 k^2_{\bot} 
+ \left( 2 \mtb{\Omega} \cdot \mtb{k} \right) ^2 }{k^2}\, , 
\label{eq:disprel}
\end{equation}
where $\mtb{k}$ is the wave vector and $k_{\bot}$ is its component
orthogonal to the apparent gravity.  These modes are generally
referred to as inertia-gravity waves by~\citet{1989nos..book.....U},
as they are restored by both gravity and the Coriolis force. In the
barotropic case, $N=0$, equation~(\ref{eq:disprel}) describes the
inertial waves, while for a non-rotating model it reduces to the usual
gravity waves.  In stratified and rotating neutron stars, it is thus
to be expected that the low-frequency oscillation modes have an hybrid
character with gravity dominating at low rotation rates and the
Coriolis force taking over at fast stellar spin. 
%
\begin{figure}
\begin{center}
\includegraphics[height=60mm]{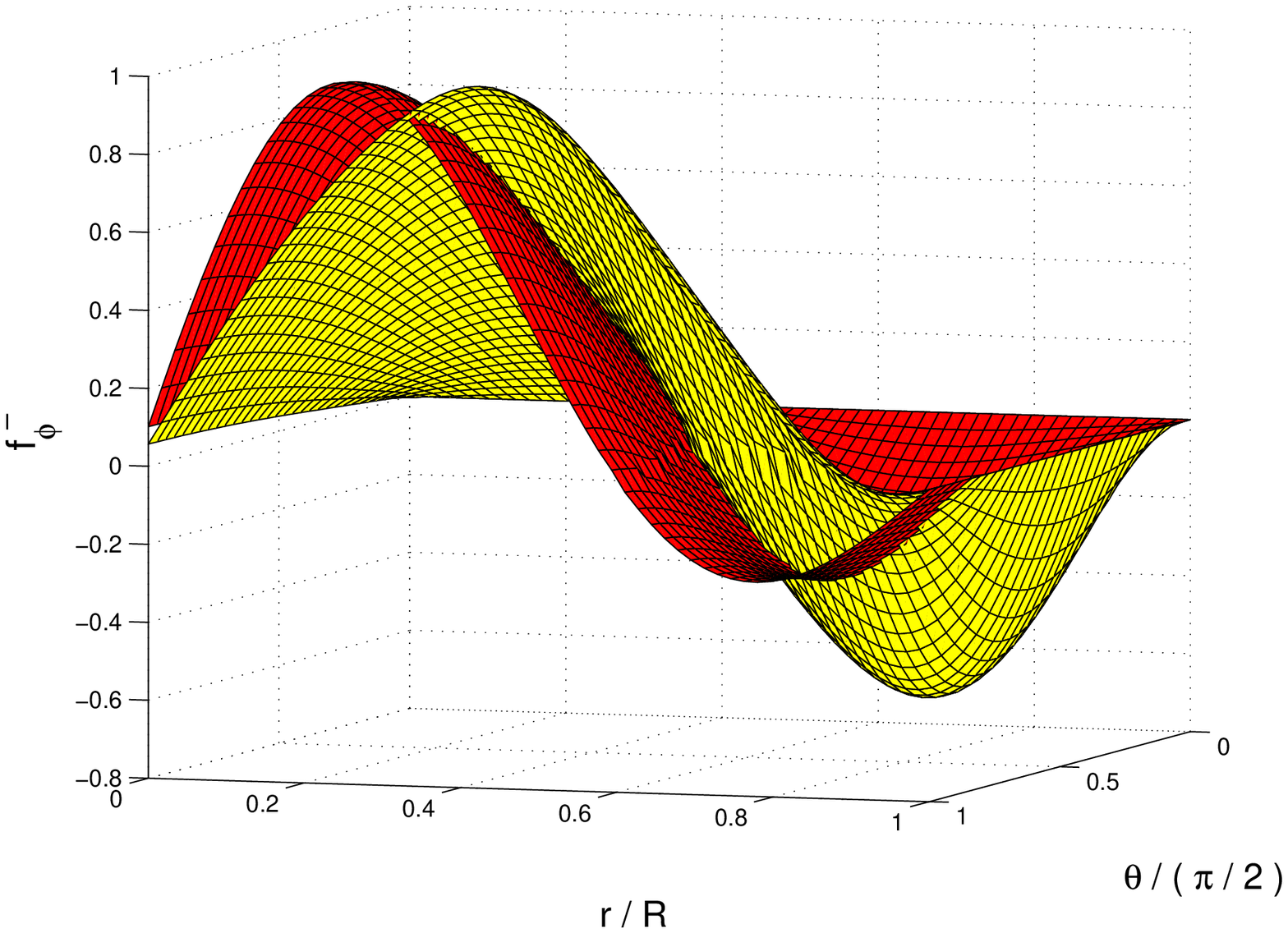}
\includegraphics[height=60mm]{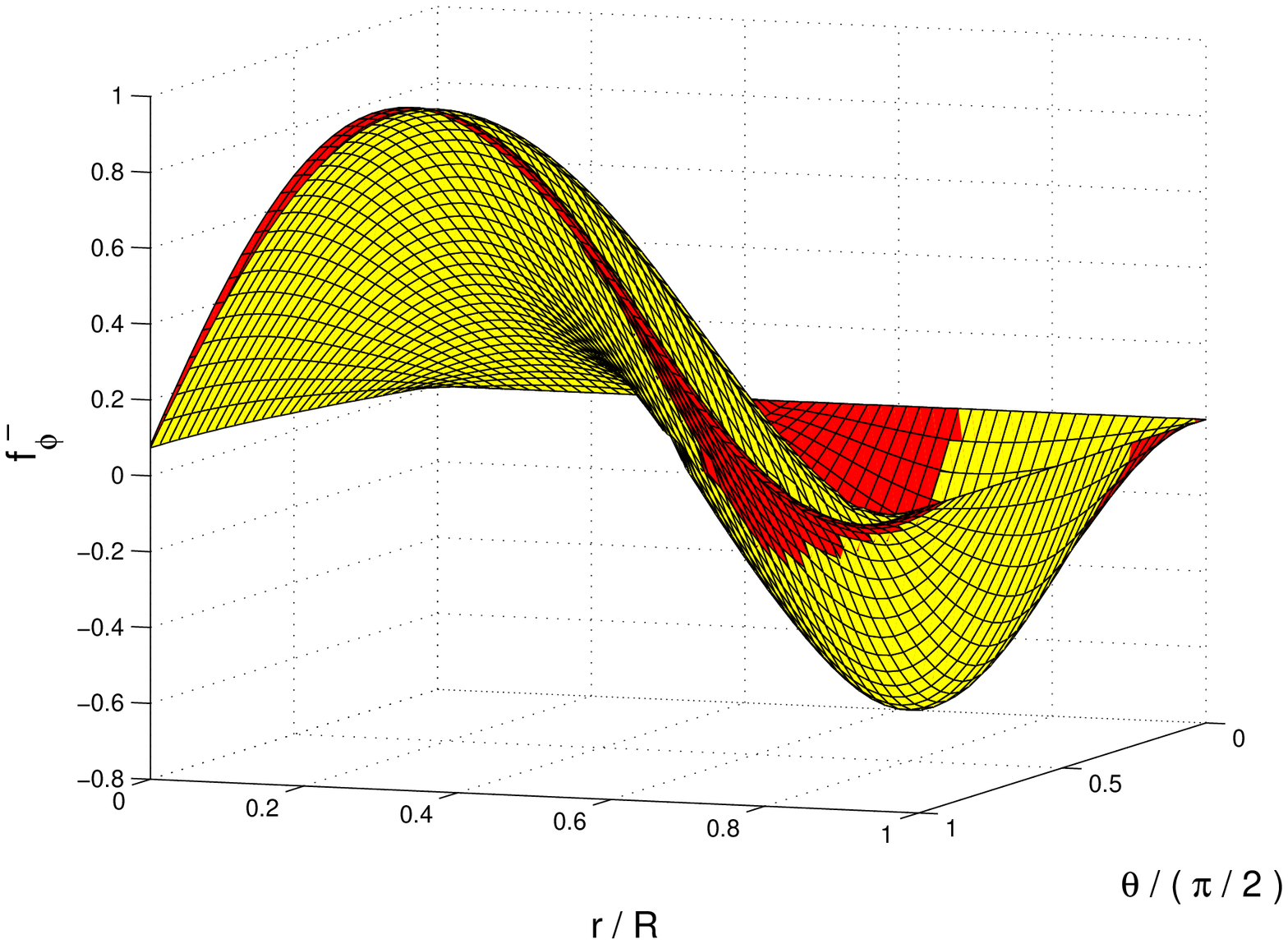}
\caption{Eigenfunctions of the flux component perturbation
 $f_{\phi}^-$ for rotating stellar models with rotation rate $\Omega /
 \sqrt{G \rho_c} = 0.084$ (left panel) and $\Omega / \sqrt{G \rho_c} =
 0.488$ (right panel).  The yellow surfaces refer to the
 ${}^3\rm{i}_{2}$ inertial mode of barotropic models with $\Gamma_\f =
 \Gamma_{\beta} = 2$, while the red surfaces represent the
 counter-rotating ${}^2\rm{g}^r_{1}$ mode of stratified stars with
 $\Gamma_\f = 2.1 $ and $ \Gamma_{\beta} = 2$.  This figure illustrates how 
 g- and inertial mode eigenfunctions become similar for
 increasing rotation rates.
 \label{fig:fphi2D}}
\end{center}
\end{figure}
\begin{figure}
\begin{center}
\includegraphics[height=55mm]{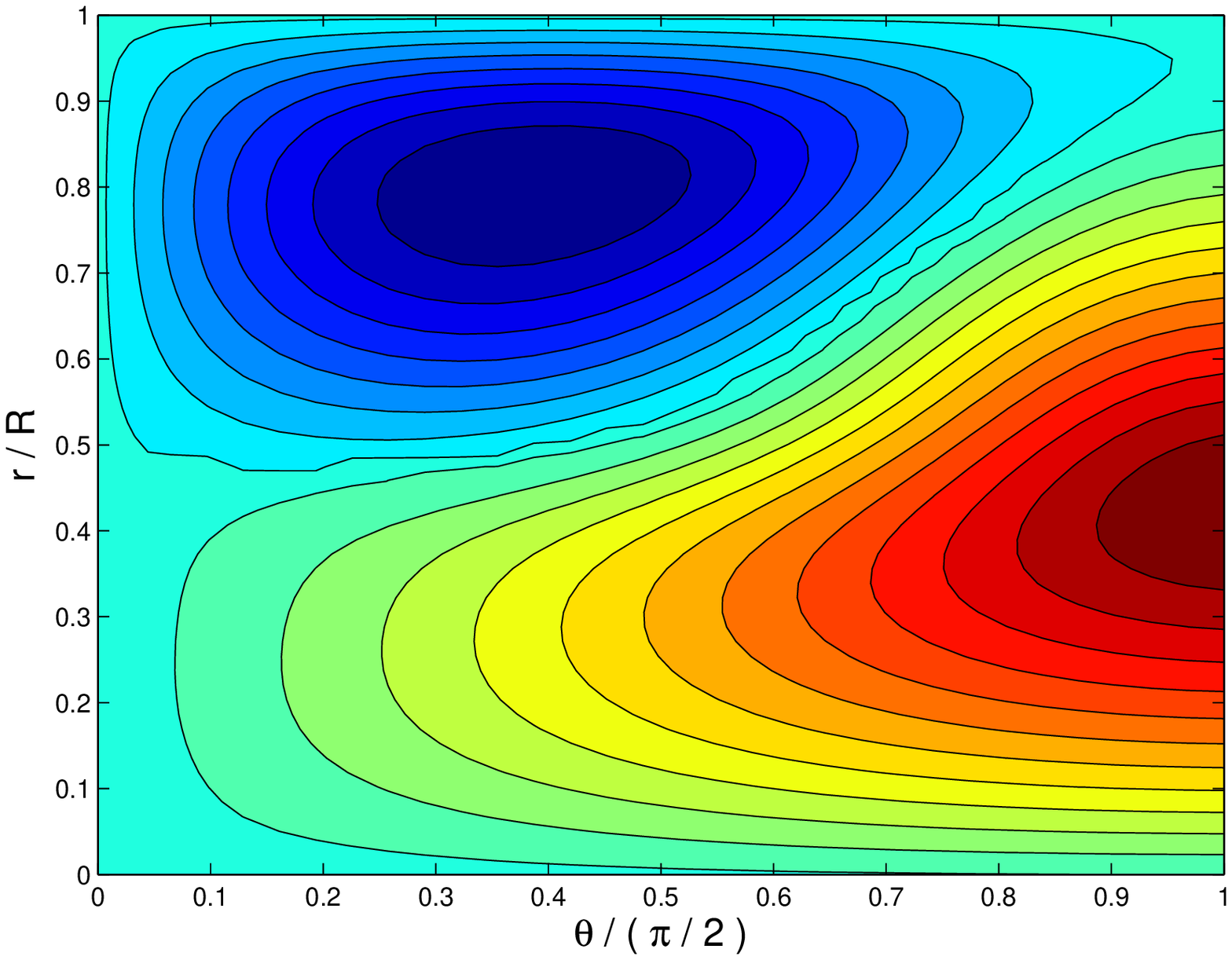}
\includegraphics[height=55mm]{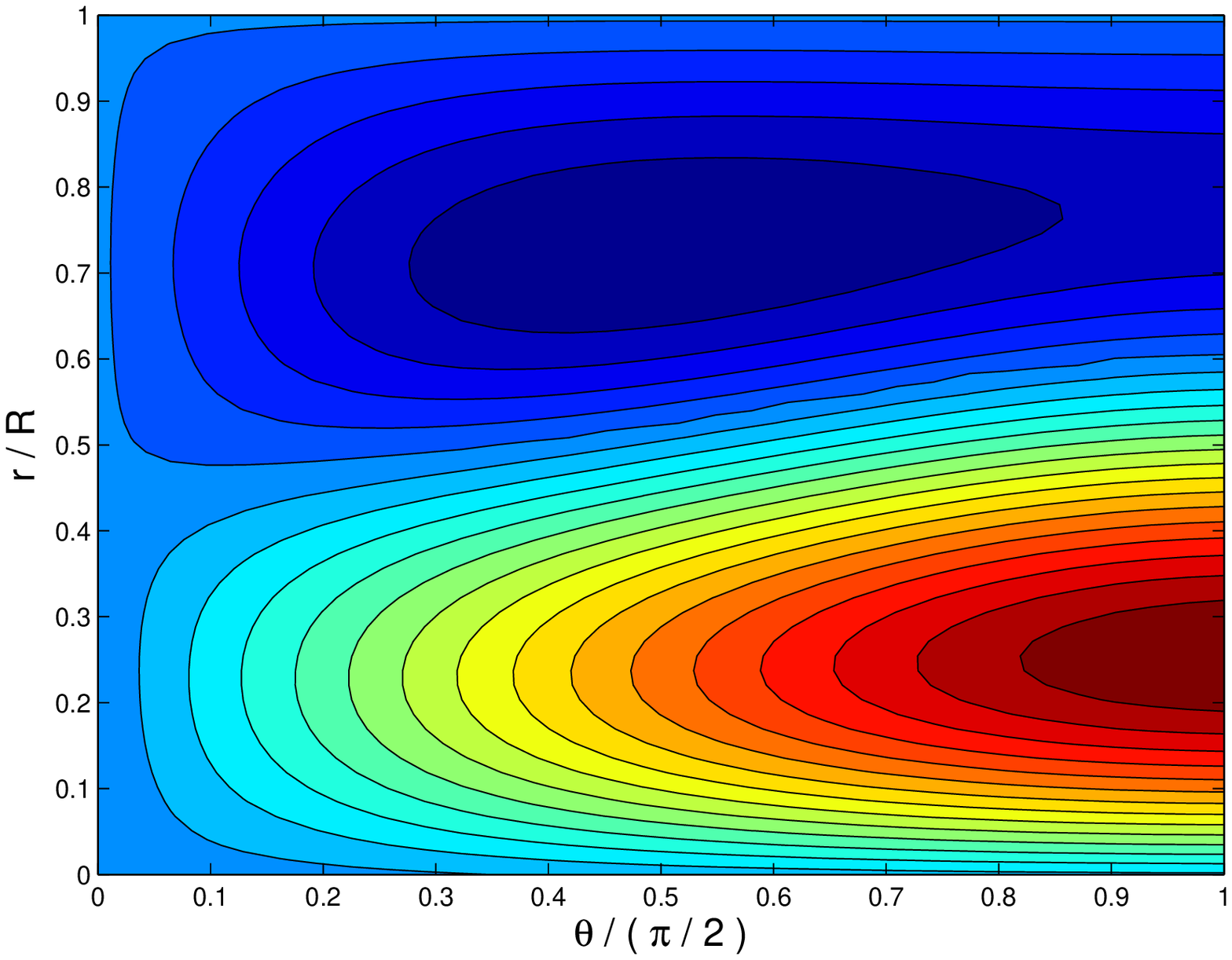}
\caption{Eigenfunctions of the flux component perturbation
 $f_{\phi}^-$ for the rotating stellar model with rotation rate
 $\Omega / \sqrt{G \rho_c} = 0.084$. The left panel shows the contour
 plot of the ${}^3\rm{i}_{2}$ inertial mode for the barotropic model
 with $\Gamma_\f = \Gamma_{\beta} = 2$.  The right panel displays the
 contour plot of the counter-rotating ${}^2\rm{g}_{1}$ mode of a
 stratified star with $\Gamma_\f = 2.1 $ and $ \Gamma_{\beta} =
 2$. These plots give a complementary view of the eigenfunctions shown
 in the left panel of Fig.~(\ref{fig:fphi2D}), and confirm the
 difference between the two modes in a slowly rotating model.
 \label{fig:fphi_contour_slow}}
\end{center}
\end{figure}
\begin{figure}
\begin{center}
\includegraphics[height=55mm]{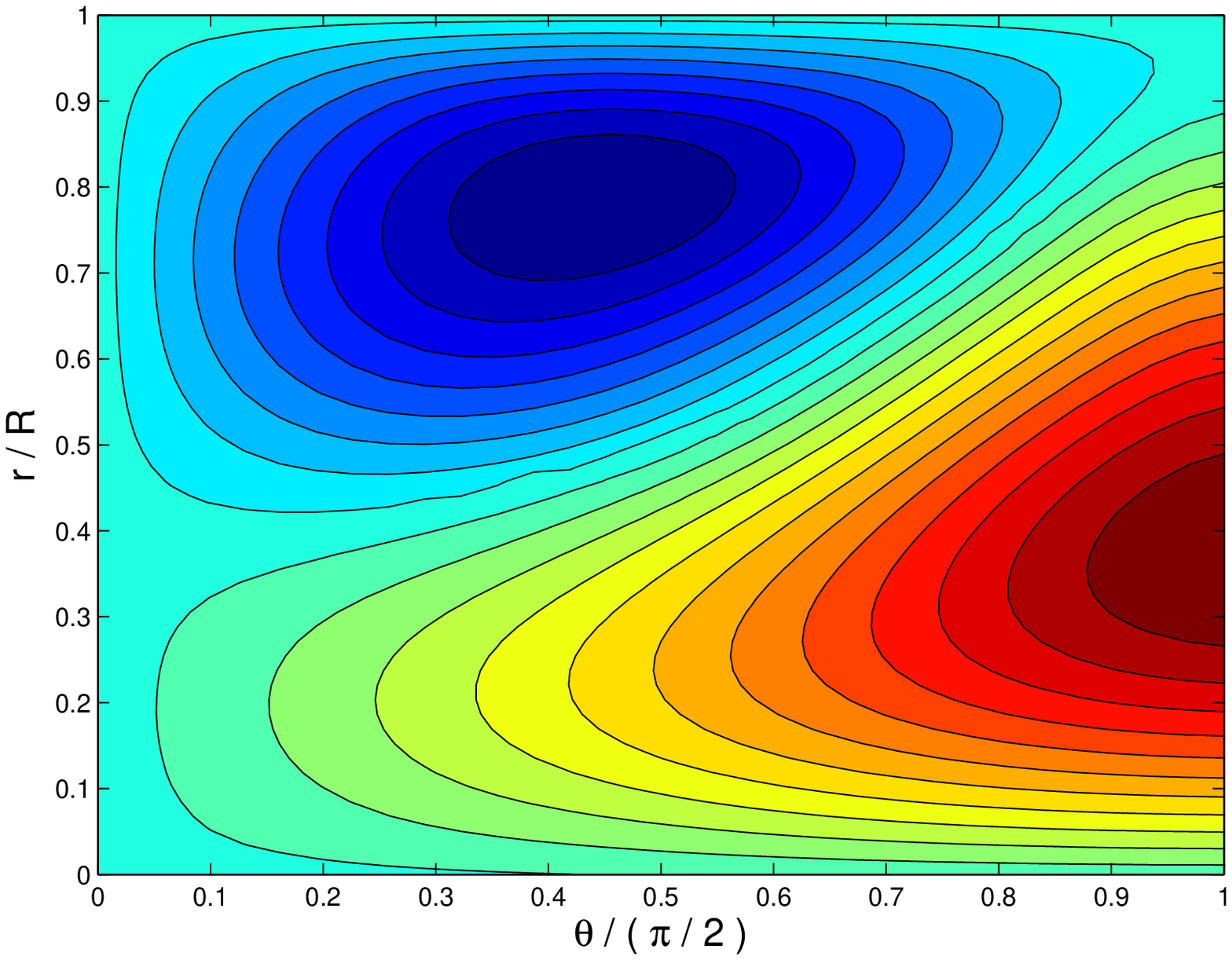}
\includegraphics[height=55mm]{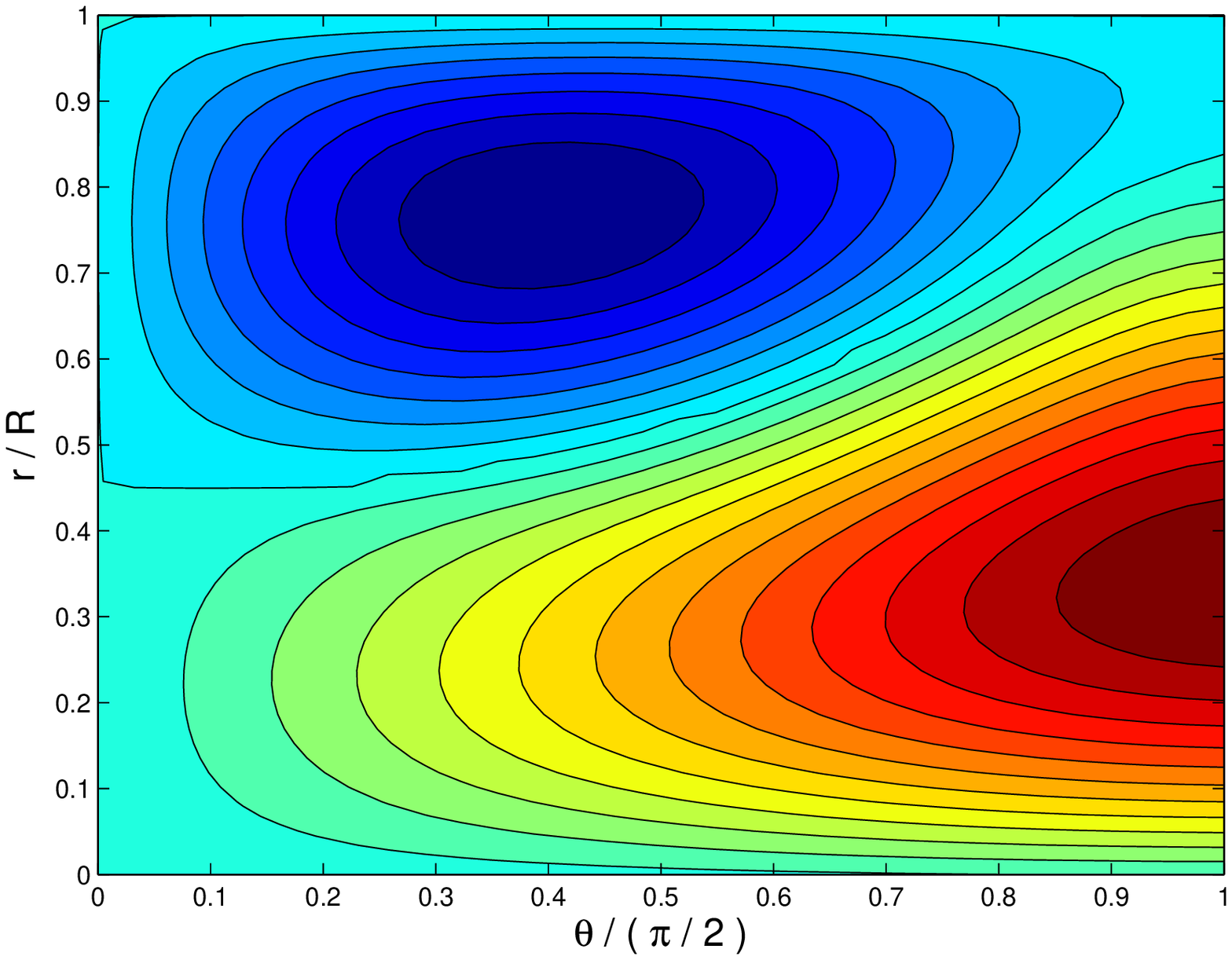}
\caption{Eigenfunctions of the flux component perturbation
 $f_{\phi}^-$ for the rotating stellar model with rotation rate
 $\Omega / \sqrt{G \rho_c} = 0.488$ The left panel shows the contour
 plot of the ${}^3\rm{i}_{2}$ inertial mode for the barotropic model
 with $\Gamma_\f = \Gamma_{\beta} = 2$.  The right panel displays the
 contour plot of the counter-rotating ${}^2\rm{g}_{1}$ mode of a
 stratified star with $\Gamma_\f = 2.1 $ and $ \Gamma_{\beta} = 2$.
 These plots confirm the results of Fig.~(\ref{fig:fphi2D}) showing
 that the structure of these two eigenfunctions becomes  similar
 in fast rotating models.
 \label{fig:fphi_contour_fast}}
\end{center}
\end{figure}

However, it is worth noting that the buoyancy and the Coriolis force
can be locally dominant in different regions of the star. This can be
understood from a propagation diagram that illustrates the spectral
properties of the star. For instance, we consider two stellar models
with frozen adiabatic index $\Gamma_\f = 2.1$ and with angular
velocity $\Omega / \sqrt{G \rho_c} = 0.084$ and $\Omega / \sqrt{G
\rho_c} = 0.488$, respectively.  For these models we determine the
Brunt-V\"{a}is\"{a}l\"{a} frequency $N$~(\ref{eq:Nfreq}), the Lamb
frequency~\citep{1989nos..book.....U}
\begin{equation}
S_l^2 = \frac{ l\left(l+1\right) }{r^2} c_s^2 \, ,
\end{equation}
and effect of the Coriolis force on the dispersion
relation~(\ref{eq:disprel}), i.e.  $2 \Omega \cos \theta$. The role of
the Lamb frequency is to estimate the propagation region for acoustic
modes.  For these two stellar models, the propagation diagram for the
$l=m=2$ modes is shown in Fig.~\ref{fig:prop-diag} using the
normalized radial coordinate $r/R$. For the slowly rotating star with
$R_p / R_{eq} = 0.996$, the $N$ and $S_l$ frequencies do not vary
much, while the effect of the Coriolis force term $2 \Omega \cos
\theta$ changes according to its angular dependence, from zero at the
equator to the maximum at the pole. In this case, the star is spinning
so slowly that the buoyance ($N$) is dominant in most of the
star. Only near the centre, for $r \lesssim 0.2 R$, is the Coriolis
force stronger. As a result, we should expect the inertia-gravity
modes to behave as g-modes. Later, this will be confirmed by a global
mode analysis.  When the star rotates faster, as in the model with
$R_p / R_{eq} = 0.85$~(right panel of Fig.~\ref{fig:prop-diag}), the
Coriolis term is more important and larger than the
Brunt-V\"{a}is\"{a}l\"{a} frequency in the most of the star. In order
to keep the figure clear, we have shown the $N$ and $S_l$ frequencies
only for $\theta = \pi /4 $. The change in the other directions is
visible, but does not modify the qualitative properties of the
propagation diagram.  However, even in weakly stratified and rapidly
rotating stars there is a region near the equator where buoyancy
dominates over the Coriolis force. This region, which is due to the
$\cos \theta$ dependence of the Coriolis term, shrinks for higher
rotation rates.

\begin{figure}
\begin{center}
\includegraphics[height=60mm]{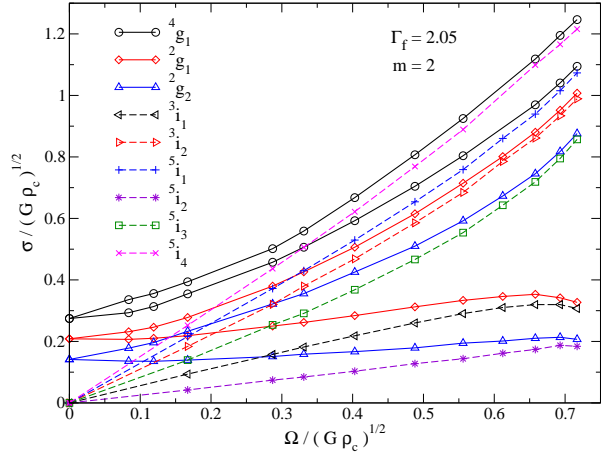}
\caption{ Frequencies of some selected g-modes (solid lines) for
$\Gamma_{\rm f}=2.05$ models and polar-led inertial modes (dashed
lines) of the barotropic $\Gamma_{\beta}=\Gamma_{\rm f}=2$ model. All
frequencies are measured in the rotating frame. The dominance of the
Coriolis force in rapidly rotating stars is evident from this
figure. We see that the
g-modes are mainly restored by the Coriolis force when $
\Omega / \sqrt{G \rho_c} > 0.3$, and thus become similar to the barotropic inertial
modes.
\label{fig:Gf2.05-g-in-comp-polar}}
\end{center}
\end{figure}

The inertia-gravity modes can be identified in our time domain
simulations by studying the FFT of the perturbation evolution and
extracting the 1D and 2D mode eigenfunctions with the code developed
by~\citet{Stergioulas:2003ep} and~\citet{Dimmelmeier:2005zk}.  In
order to study the dependence of the modes on the stellar spin and
stratification, we construct a set of equilibrium models by gradually
varying the two parameters $\Omega$ and $\Gamma_{\rm f}$.  We can then
follow the change of the mode frequency and eigenfunction shape and
correctly distinguish it among the other peaks in the spectrum.  Let
us consider, for instance, a sequence of slowly rotating models with
fixed rotation rate $\Omega / \sqrt{G\rho_c} = 0.167$ and different
$\Gamma_{\rm f}$.  Typical FFT results are shown in Fig.~\ref{fig:Sp},
where we have identified the ${}^3\rm{i}_{2}$ inertial mode for the
barotropic model and the counter-rotating ${}^2\rm{g}^r_1$ mode for a
set of stars with increasing composition gradients.  The results
suggest that the g-mode frequency approaches the barotropic inertial
mode in the limit $\Gamma_{\rm f}\to 2$.  This association becomes
more evident if we combine these results for a set of rotating models,
see Fig.~\ref{fig:g1r-i31}.  When the star rotates with $ 2 \Omega
\gtrsim N$, the retrograde ${}^2\rm{g}^r_1$ mode seems to behave as
the barotropic ${}^3\rm{i}_{2}$ inertial mode.  This transition
obviously depends on the level of stratification. This property is
also noted in the mode eigenfunctions.  In
Figs.~\ref{fig:fphi2D}--\ref{fig:fphi_contour_fast}, we show how the
eigenfunctions of the mass flux component $f_{\phi}^{-}$ of the
counter-rotating ${}^2\rm{g}^r_1$ mode approach the shape of the
${}^3\rm{i}_{2}$ mode when $ 2 \Omega > N$. In this case, the
non-barotropic star has $\Gamma_{\rm f} = 2.1$ and an estimated
Brunt-V\"{a}is\"{a}l\"{a} frequency $\langle N / \sqrt{G \rho_c}
\rangle \simeq 0.447$, see equation~(\ref{NN}).  In
Fig.~\ref{fig:fphi2D} we provide a three-dimensional picture of the
eigenfunctions, for rotation rate $\Omega / \sqrt{G \rho_c} = 0.084$
in the left panel, and for $\Omega / \sqrt{G \rho_c} = 0.488$ in the
right panel (the latter corresponding to a star with an angular
velocity slightly larger than the Brunt-V\"{a}is\"{a}l\"{a}
frequency).  Clearly, for the slower rotation rate (left panel) the
$f_{\phi}^{-}$ the ${}^2\rm{g}_1$ eigenfunction is quite different
from the eigenfunction of the barotropic ${}^3\rm{i}_{2}$ mode. In
particular, the nodal curve of the two modes do not coincide near the
equator.  In contrast for the more rapidly rotating star (right panel)
the nodal curves of the two modes are very similar.  This can be seen
particularly clearly in the contour plots of the eigenfunctions given
in Fig.~\ref{fig:fphi_contour_slow} (for $\Omega / \sqrt{G \rho_c} =
0.084$) and Fig.~\ref{fig:fphi_contour_fast} (for $\Omega / \sqrt{G
\rho_c} = 0.488$).

\begin{figure}
\begin{center}
\includegraphics[height=60mm]{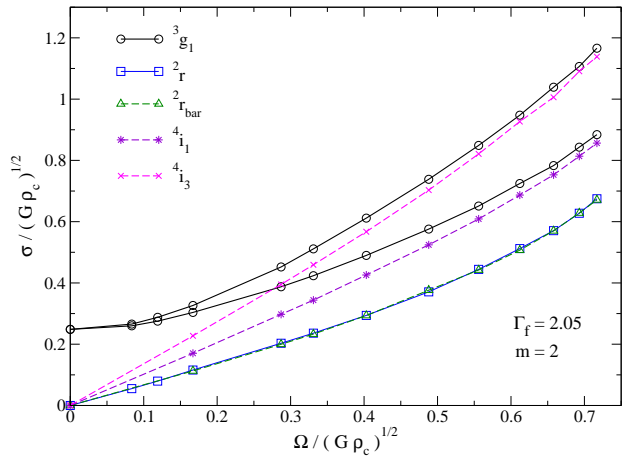}
\caption{Frequencies of some selected $m=2$ g-modes (solid lines) for
$\Gamma_{\rm f}=2.05$ models and axial-led inertial modes (dashed
line) of the barotropic models $\Gamma_{\beta}=\Gamma_{\rm
f}=2$. Frequencies are measured in the rotating frame. The counter-
and co-rotating ${}^3\rm{g}_{1}$ modes approach the frequencies of the
axial-led inertial modes in the limit of fast rotation. In contrast,
the r-mode is essentially unaffected by composition gradients.
\label{fig:Gf2.05-g-in-comp-axial}}
\end{center}
\end{figure}

When we extend this approach to the other g and inertial modes, we
find further mode associations.  Combined results are provided in
Figs.~\ref{fig:Gf2.05-g-in-comp-polar}
and~\ref{fig:Gf2.05-g-in-comp-axial}. These figures display the
spectrum of barotropic models and stratified stars with $\Gamma_{\rm
f} = 2.05$.  The results show that individual non-barotropic g-modes
tend towards specific barotropic inertial modes as the rotation rate
of the star is increased. This agrees with the notion that the
Coriolis force dominates the buoyancy in fast spinning systems. From
the results in the two figures we can identify various g-modes with
barotropic inertial-mode counterparts. For type I parity and $m=2$
this leads to: \\
\begin{equation}
{}^4 \rm{g}_{1}^{r} \longleftrightarrow {}^5 \rm{i}_4 , \qquad {}^4 \rm{g}_{1}^{p} \longleftrightarrow {}^5 \rm{i}_1  ,
\qquad {}^2 \rm{g}_{1}^{r} \longleftrightarrow {}^3 \rm{i}_2 ,\qquad  {}^2 \rm{g}_{1}^{p} \longleftrightarrow {}^3 \rm{i}_1  ,
\qquad {}^2 \rm{g}_{2}^{r} \longleftrightarrow {}^5 \rm{i}_3  , \qquad {}^2 \rm{g}_{2}^{p} \longleftrightarrow {}^5 \rm{i}_2 \, ,
\label{eq:connrule1} \end{equation}
where the upper index r and p denotes retrograde and prograde g-modes
respectively.  Meanwhile, for type II parity and $m=2$ we find that:
\\
\begin{equation}
{}^3 \rm{g}_{1}^{r} \longleftrightarrow {}^4 \rm{i}_3 \, , \qquad {}^3
\rm{g}_{1}^{p} \longleftrightarrow {}^4 \rm{i}_1 \, . \label{eq:connrule2}
\end{equation}
From Fig.~\ref{fig:Gf2.05-g-in-comp-axial} we also see that it is only
the $l=m$ r-mode that does not have a g-mode counterpart. As expected,
the r-mode is essentially the same in the barotropic and
non-barotropic models. In the type II parity perturbations we could
accurately determine only the first g-mode, as the r-mode peak was
dominant in the FFT and overwhelmed the higher order g-modes.

These results are consistent with the study
of~\citet{2000ApJS..129..353Y}, where the authors worked within the
slow rotation approximation and in the frequency domain. They studied
inertia-gravity modes of $\Gamma_\beta = 2$ polytropic models with
weak stratification and slow rotation. In fact, their frozen adiabatic
index corresponds to $\Gamma_{\f} = 2.0004$ and the fastest rotating
model they consider represents $\Omega / (G M /R^3 )^{1/2} = 0.1$,
where $M$ and $R$ are respectively the mass and radius of the
non-rotating models.  In our units this rotation rate in equivalent to
$\Omega / ( G \rho_c ) ^{1/2} = 0.113$ and an axis ratio $R_p / R_{eq}
\gtrsim 0.992$. In contrast, in the present work we investigate
rapidly rotating models up to $\Omega / ( G \rho_c ) ^{1/2} = 0.717$
($R_p / R_{eq} = 0.6$) and with larger composition gradients.  We have
tested our results against those of~\citet{2000ApJS..129..353Y} by
numerically evolving $m=2$ perturbations of the two slowest rotating
models reported in Table~\ref{tab:back-models} with $\Gamma_{\f} =
2.0004$. The frequencies have an accuracy to better than 2
percent. Finally, equations~(\ref{eq:connrule1})
and~(\ref{eq:connrule2}) respect the connection rules reported
in~\citet{2000ApJS..129..353Y}.

\section{Concluding remarks\label{conclusions}} \label{sec:concl}

Information concerning the oscillation spectra of realistic neutron
star models helps develop our understanding of the physics required to
describe these objects. Any spectral property can in principle be
attributed to particular physical quantities or configurations of the
star. Astero-seismology studies, using either electromagnetic or
gravitational signals (or indeed both) may thus help constrain the
state of matter at supernuclear densities.  Of course, in order to
facilitate this, we need to improve our theoretical models and clarify
the origin of various stellar pulsation features.

In this paper, we have studied the pulsations of stratified and
rapidly rotating neutron stars with the aim of understanding the
dependence of the composition g-modes on the rotation rate of the
star.  The stellar pulsations were studied using the linearised Euler
and conservation equations on an axisymmetric background. In order to
simplify the problem, we used the Cowling approximation where the
gravitational potential perturbations are neglected. This
approximation generates only a small error in the g-mode frequencies.
The system of perturbation equations was evolved in time by a 2D
numerical code based on a standard finite differencing scheme. This
code was tested against various results available in literature and
demonstrated good accuracy.

Since both Coriolis and buoyancy forces act on a perturbed fluid
element of a rotating and stratified star, the low-frequency modes
have a mixed inertia-gravity character. They typically behave as
g-modes in the slowly rotating limit, while they assume the properties
of the inertial modes when the star rotates rapidly.  By comparing the
oscillation frequencies and the associated eigenfunctions of
barotropic inertial modes and the inertia-gravity modes of
non-barotropic stars, we have demonstrated how the two sets of modes
are associated at fast rotation rates.

Our results show that it would be difficult to deduce the presence of
composition gradients from the inertia-gravity mode spectrum of fast
rotating stars.  However, as the neutron star ages and spins down, the
dependence of the inertia-gravity modes on the star's rotation and
their deviation from the inertial barotropic modes can, at least in
principle, be used to estimate the Brunt-V\"{a}is\"{a}l\"{a} frequency
and the degree of stratification.

As an extension of this work, we are currently developing a numerical
code for studying the dynamics of rapidly rotating superfluid neutron
stars. This is an important development since all mature neutron stars
are expected to have superfluid components in the core. It is also
well known that the associated multi-fluid dynamics leaves an imprint
on the stellar oscillation spectrum, see \citet{2007arXiv0709.0660L}
for a recent discussion.  We hope to be able to report on the initial
results of this investigation soon.

\section*{Acknowledgements}
This work was supported by STFC through grant number PP/E001025/1. NA
also acknowledges support from STFC via Senior Research Fellowship no
PP/C505791/1.


\nocite*

\label{lastpage}
\end{document}